 \documentclass[preprint2]{aastex}

\usepackage{float}
\slugcomment{Accepted by Astronomical Journal}

\shorttitle{CHANG-ES IV: Radio continuum emission of 35 edge-on galaxies}
\shortauthors{Wiegert et al.}

\begin{document}


\title{CHANG-ES IV: Radio continuum emission of 35 edge-on galaxies observed with the Karl G. Jansky
Very Large Array in D-configuration -- Data Release 1}


\author{Theresa Wiegert\altaffilmark{1}, Judith Irwin\altaffilmark{2}, Arpad Miskolczi\altaffilmark{3}, Philip Schmidt\altaffilmark{4}, Silvia Carolina Mora\altaffilmark{5}, Ancor Damas-Segovia\altaffilmark{6}, Yelena Stein\altaffilmark{7}, Jayanne English\altaffilmark{8}, Richard J. Rand\altaffilmark{9}, Isaiah Santistevan\altaffilmark{10}, Rene Walterbos\altaffilmark{11}, Marita Krause\altaffilmark{12}, Rainer Beck\altaffilmark{13}, Ralf-J{\"u}rgen Dettmar\altaffilmark{14}, Amanda Kepley\altaffilmark{15}, Marek Wezgowiec\altaffilmark{16}, Q. Daniel Wang\altaffilmark{17}, George Heald\altaffilmark{18},  Jiangtao Li\altaffilmark{19}, Stephen MacGregor\altaffilmark{20}, Megan Johnson\altaffilmark{21}, A. W. Strong\altaffilmark{22}, Amanda DeSouza\altaffilmark{23}, Troy A. Porter\altaffilmark{24}} 

\altaffiltext{1}{Dept. of Physics, Engineering physics \& Astronomy, Queen's University,
    Kingston, ON, Canada, K7L 3N6; {\tt twiegert@astro.queensu.ca}}
\altaffiltext{2}{Dept. of Physics, Engineering Physics \& Astronomy, 
Queen's University, Kingston, ON, Canada, K7L 3N6, {\tt irwin@astro.queensu.ca}.}
\altaffiltext{3}{Astronomisches Institut, Ruhr-Universit{\"a}t Bochum, 44780 Bochum, Germany, 
{\tt miskolczi@astro.rub.de}.}
\altaffiltext{4}{Max-Planck-Institut f{\"u}r Radioastronomie,  Auf dem H{\"u}gel 69,
53121, Bonn, Germany,
{\tt pschmidt@mpifr-bonn.mpg.de}.}
\altaffiltext{5}{Max-Planck-Institut f{\"u}r Radioastronomie,  Auf dem H{\"u}gel 69,
53121, Bonn, Germany,
{\tt cmora@mpifr-bonn.mpg.de}.} 
\altaffiltext{6}{Max-Planck-Institut f{\"u}r Radioastronomie,  Auf dem H{\"u}gel 69,
53121, Bonn, Germany,
{\tt adamas@mpifr-bonn.mpg.de}.} 
\altaffiltext{7}{Astronomisches Institut, Ruhr-Universit{\"a}t Bochum, 44780 Bochum, Germany, 
{\tt yelena.stein@astro.ruhr-uni-bochum.de}.}
\altaffiltext{8}{Department of Physics and Astronomy, 
University of Manitoba, Winnipeg, Manitoba, Canada, R3T 2N2,
{\tt jayanne\_english@umanitoba.ca}.}
\altaffiltext{9}{Dept. of Physics and Astronomy, University of New Mexico, 
800 Yale Boulevard, NE, Albuquerque, NM, 87131, USA, {\tt rjr@phys.unm.edu}.} 
\altaffiltext{10}{Dept. of Physics and Astronomy, University of New Mexico, 
800 Yale Boulevard, NE, Albuquerque, NM, 87131, USA, {\tt isantistevan@unm.edu}.} 
\altaffiltext{11}{Dept. of Astronomy, New Mexico State University, 
PO Box 30001, MSC 4500, Las Cruces, NM 88003, USA, {\tt rwalterb@nmsu.edu}.}
\altaffiltext{12}{Max-Planck-Institut f{\"u}r Radioastronomie,  Auf dem H{\"u}gel 69,
53121, Bonn, Germany,
{\tt mkrause@mpifr-bonn.mpg.de}.} 
\altaffiltext{13}{Max-Planck-Institut f{\"u}r Radioastronomie, Auf dem H{\"u}gel 69,
53121, Bonn, Germany,
{\tt rbeck@mpifr-bonn.mpg.de}.}
\altaffiltext{14}{Astronomisches Institut, Ruhr-Universit{\"a}t Bochum, 44780 Bochum,
 Germany,
{\tt dettmar@astro.rub.de}.}
\altaffiltext{15}{NRAO Charlottesville, 520 Edgemont Road, Charlottesville, VA 22903-2475, USA, 
{\tt akepley@nrao.edu}.} 
\altaffiltext{16}{Astronomisches Institut, Ruhr-Universit{\"a}t Bochum, 44780 Bochum, Germany, {\tt markmet@oa.uj.edu.pl}.}
\altaffiltext{17}{Dept. of Astronomy, University of Massachusetts, 710 North
Pleasant St., Amherst, MA, 01003, USA, 
{\tt wqd@astro.umass.edu}.}
\altaffiltext{18}{Netherlands Institute for Radio Astronomy (ASTRON), 
Postbus 2, 7990 AA, Dwingeloo, The Netherlands,
{\tt heald@astron.nl}.}
\altaffiltext{19}{Dept. of Astronomy, University of Michigan, 311 West Hall, 1085 S. University Ave., Ann Arbor, MI 48109, USA, 
{\tt jiangtal@umich.edu}.} 
\altaffiltext{20}{Dept. of Physics, Engineering physics \& Astronomy, Queen's University,
    Kingston, ON, Canada, K7L 3N6; {\tt 11sm36@queensu.ca}}
\altaffiltext{21}{CSIRO, PO Box 76, Epping, NSW 1710, Australia, 
{\tt megan.johnson@csiro.au}.}
\altaffiltext{22}{Max-Planck-Institut f{\"u}r extraterrestrische Physik, 
Garching bei M{\"u}nchen, Germany, {\tt aws@mpe.mpg.de}.}
\altaffiltext{23}{Dept. of Physics, Engineering physics \& Astronomy, Queen's University,
    Kingston, ON, Canada, K7L 3N6; {\tt adesouza@astro.queensu.ca}}
\altaffiltext{24}{Hansen Experimental Physics Laboratory, Stanford University, 
452 Lomita Mall, Stanford, CA, 94305, USA, {\tt tporter@stanford.edu}.}

%\and

%\author{Last name}
%\affil{Affiliation}


%\altaffiltext{1}{Visiting Astronomer, Cerro Tololo Inter-American Observatory.
%CTIO is operated by AURA, Inc.\ under contract to the National Science
%\altaffiltext{2}{Society of Fellows, Harvard University.}
%\altaffiltext{3}{present address: Center for Astrophysics,
%    60 Garden Street, Cambridge, MA 02138}
%\altaffiltext{4}{Visiting Programmer, Space Telescope Science Institute}
%\altaffiltext{5}{Patron, Alonso's Bar and Grill}


\begin{abstract}
We present the first part of the observations made for the Continuum Halos in Nearby Galaxies, an EVLA Survey (CHANG-ES) project. The aim of the CHANG-ES project is to study and characterize the nature of radio halos, their prevalence, as well as their magnetic fields and the cosmic rays illuminating these fields. This paper reports observations with the compact D-configuration of the Karl G. Jansky Very Large Array (VLA) for the sample of 35 nearby edge-on galaxies of CHANG-ES. With the new wide bandwidth capabilities of the VLA, an unprecedented sensitivity was achieved for all polarization products. The beam resolution is an average of 9.6\arcsec\ and 36\arcsec\, with noise levels reaching approximately 6 and 30 $\mu$Jy/beam for C and L bands respectively (robust weighting).

We present intensity maps in these two frequency bands (C and L), with different weightings, as well as spectral index maps%with weighted spectral index values
, polarization maps and new measurements of star formation rates. The data products described herein are available to the public in CHANG-ES Data Release available at {\tt http://www.queensu.ca/changes}.

We also present evidence of a trend between galaxies with larger halos having higher SFR surface density, and show,
for the first time, a radio continuum image of the median galaxy, taking advantage of the collective signal-to-noise of 30 of our galaxies. This image shows clearly that a 'typical' spiral galaxy is surrounded by a halo of magnetic fields and cosmic rays. 
\end{abstract}




\section{Introduction}
This is the fourth paper in the series "Continuum Halos in Nearby Galaxies - an EVLA Survey" \citep[CHANG-ES,][]{IrwinI}.  The overall aims of the CHANG-ES project are to investigate the occurrence, origin and nature of radio halos, to probe the disk-halo interface, and to investigate in-disk emission. We seek to understand connections between radio halos and the host disk and its environment, and to investigate the magnetic fields in these galaxies and their halos. This covers a wide variety of science, including cosmic ray transport and wind speeds, the nature and origin of galactic magnetic fields, the far-infrared radiation - radio continuum correlation, cosmic rays and high-energy modelling, disk star formation rates and presence or absence of active galactic nuclei (AGN). Note that in this context, we use the word "halo" to refer to gas, dust, cosmic rays and the magnetic field above and below the galaxy disk, not to be confused with stellar or dark matter halos. Specifically, we call emission on larger scales, i.e. scale height $z>1$ kpc, halo emission, while the disk-halo interface is at $0.2<z<1$ kpc \citep[as defined in ][]{IrwinI}. 

Magnetic fields in the halos of edge-on galaxies, and their commonly X-shaped behaviour (magnetic field lines outside the projected galaxy's disk are bending away from the disk with increasing vertical components) are discussed in the review by \citet{Krause2009} (also see references therein). Their intensities and degree of uniformity can be estimated from the total and polarized synchrotron emission, but many uncertainties pertaining to their structures and origins in external galaxies remain \citep{HaverkornHeesen2012}.

With CHANG-ES, we have observed thirty-five nearby edge-on galaxies in the radio continuum in L and C bands (centred at approximately 1.5 and 6 GHz, respectively), in three array configurations (B, C, D; in B-configuration only L-band was observed) of the Karl G. Jansky Very Large Array (hereafter VLA). The recently enhanced VLA allows us to trace radio continuum emission at levels fainter than previously possible via its wide bandwidth capabilities. Moreover, observing on a variety of angular scales, the VLA provides distinct advantages in understanding disk-halo and halo features. For example, both faint diffuse emission and distinct filamentary structures can be investigated through the combined use of compact and extended configurations. The low frequencies, chosen because of their sensitivity to synchrotron emission, in combination with observing all polarization products, enables us to derive information about the halo magnetic fields and cosmic rays. 

Our thirty-five edge-on galaxies have inclinations higher than 75 degrees. They also adhere to limitations on declination (more than $-23$ degrees in order to be observed with sufficient uv coverage with the VLA), as well as optical diameter (4\arcmin $< d_{25} <$ 15\arcmin) and flux density ($S_{1.4GHz} \geq$ 23 mJy). Three galaxies (NGC~5775, NGC~4565 and NGC~4244) just outside of these criteria were included in the sample as well, due to evidence for extra-planar gas and availability of good ancillary data. We refer to Table 1 of \citet{IrwinI} (Paper I) for details of the galaxy sample. 

The full project, with its motivation and science goals, is presented in detail in \citet{IrwinI} (Paper I). 
Two other papers, \citet{IrwinII} (Paper II) and \citet{IrwinIII} (Paper III), present detailed results of CHANG-ES observations of NGC~4631 and UGC~10288, respectively. 

In this, the fourth CHANG-ES paper, we present {\it all} observations which were carried out in the shortest baseline array configuration, D, and display each galaxy of the survey with its results in the appendix. In particular, we show the Stokes I maps and spectral index maps for each of the galaxies in the two frequency bands, as well as the polarization map with apparent B-vectors superposed, as derived from the Stokes Q and U maps. These data products are available for download at {\tt http://www.queensu.ca/changes}.

Additionally, %weighted spectral indices, 
new star formation rates and flux densities are presented. 

The paper is organized as follows: in Section~2 we give a description of the sample selection, the setup of the observations and observation details. Section~3 describes the data reduction with calibration procedures and Section~4 presents the resulting data products, which are displayed in the appendix.
Analysis is presented in section~5 and the conclusions can be found in Section~6.

%%%%%%%%%%%%%%%%%%%%%%%%%%%%%%%%%%%%%%%%%%%%%%%%%%%%%%%%%%%%%%%%%%%%%%%%%%%%%

\section{Observation setup}
\subsection{Survey design}
The details of the observations are presented in Table~\ref{tab:observations}. 
The observations in C-band (central frequency 6.000 GHz) cover a bandwidth of 2 GHz (4.979 -- 7.021 GHz), in 16 spectral windows and 1024 spectral channels (64 in each spectral window). The L-band (central frequency 1.575 GHz) observations cover a bandwidth of 512 MHz (1.247 -- 1.503 GHz, 1.647 -- 1.903 GHz), in 32 spectral windows, and 2048 spectral channels. In L-band, we placed the two base bands, of 16 spectral windows each, 144 MHz apart in order to avoid a region of particularly strong and contaminating radio frequency interference (RFI). Note that due to flagging during the reductions, the final central frequencies shown in Tables~\ref{tab:imagingparametersC} and \ref{tab:imagingparametersL} for each galaxy are adjusted from the values given here. 

\subsubsection{Sensitivity}
The theoretical noise level is approximately 6 $\mu$Jy/beam in C-band (with confusion limit and flagging taken into account). Due to a high uncertainty in confusion, the theoretical estimate for L-band of 89 $\mu$Jy/beam is substantially higher than the actual value we achieved, which is around 30-35 $\mu$Jy/beam. However, due to variable noise in the L-band, we measured the noise in areas far from the source where it was more uniform (see Sec.~\ref{sec:rmsnoise}). Additionally, a few galaxies exhibit higher values for reasons outlined later in this paper. 

\subsubsection{Distances}
\label{sec:distances}
We have made a few modifications to the distances used in Table 1 of Paper I. Five galaxies have had their distances modified, whereas the rest remain the same as listed in Paper I. 

The five galaxies with modified distances are: NGC~891, NGC~4244, NGC~4565, NGC~4631, and NGC~5907. These were adjusted because they have distances derived from the Tip of the Red Giant Branch (TRGB) \citep{RadburnSmithetal2011}.
Because of the reliability of the TRGB method, and that it likely provides the best distances for edge-on systems, we have adopted those distances here. The most significant distance change  was for NGC~4565, which went from a Hubble Flow distance of 27.1 Mpc to a TRGB distance of 11.9 Mpc. For the other four galaxies the changes were minor (see Table~\ref{tab:observations}).

Of the remaining galaxies, the distances of the five Virgo cluster galaxies (denoted with V in Table~\ref{tab:observations}) were adopted from \citet{Solanesetal2002}. 

All other galaxy distances were taken from NED, using the "Hubble flow corrected for Virgo and the Great Attractor" (HF) distance, and a Hubble constant of 73 km/s/Mpc. Since a Hubble flow method may not always be accurate for nearby galaxies at low heliocentric velocities, we reviewed the galaxy distances for this paper. We compared the HF distance with the median of various Tully-Fisher (TF) derived distances. The latter are listed in NED as "red-shift independent" distance derivations. Note that we excluded the "Tully 1988" values listed in NED since those are Hubble Flow based. For the majority of the galaxies, the changes between the two methods were well within the uncertainties associated with either
method. In particular, it is not uncommon to find a factor 2 difference in the various TF-method published distances. We therefore retained the distances used in Paper I for these 25 systems. The comparison showed that for 16 galaxies the agreement between the adopted HF distance and the median TF distances was better than 25\%, for 9 galaxies the difference was larger than 25\%. The galaxies with the largest distance uncertainty ($>$ 35\%) include NGC~2683, NGC~3432, NGC~4666, NGC~5775, and NGC~5792. 
See Table~\ref{tab:observations} for our current distance list. 

%%%%%%%%%%%%%%%%%%%%%%%%%%%%%%%%%%%%%%%%%%%%%%%%%%%%%%%%%%%%%%%%%%%%%%%%%%%%%

\subsection{Observations}
Of the 405 hours which were awarded for the entire CHANG-ES project, 65 hours were set apart for the D-configuration observations in two frequency bands, L and C. The observations were divided up into 13 scheduling blocks, each of which contained scans of one primary gain and phase calibrator (hereafter referred to as the primary calibrator) and one zero polarization calibrator to calibrate polarization leakage from the instrumentation. Additionally, complex gain calibrations were performed using a source (hereafter the secondary calibrator) less than 10 degrees from the target galaxy. 
The secondary calibrator was observed before and after the galaxy scans, every 20-25 minutes. The bulk of the data was observed during December 2011. Two scheduling blocks were re-observed in March 2013. 
%{\bf [mention the reasons? not necessary, but keeping here in case needed: Group1-L2 due to wrong start time, shadowing for low dec gal, Group 10 unusually large fraction had to be flagged)}. 
See Table~\ref{tab:observations} for observation specifics.
One scheduling block observed in C-band, ID 5062559, suffered increased antenna system temperatures at times coinciding with bad weather (snow). Consequently, the rms noises for galaxies NGC~4302, NGC~4388, NGC~4438, NGC~4594, NGC~4666 and NGC~4845 increased to approximately twice the theoretical expectation at C-band. 

\subsubsection{Large galaxies}
\label{sec:twopointings}
Eight galaxies in the sample are too large to fit inside the primary beam of 7.5\arcmin\ FWHM at C-band. We thus observed these in two pointings, separated by half the diameter of the primary beam at half maximum, i.e. one quarter beam from the galaxy centre along the disk on either side of the centre. The two-pointing galaxies are marked with an asterisk beside their names in Table~\ref{tab:observations}. Table~\ref{tab:twopointings} lists the coordinates for the separate pointings. One single pointing was sufficient for all galaixes in L-band, where the diameter of the primary beam is 30\arcmin.

%%%%%%%%%%%%%%%%%%%%%%%%%%%%%%%%%%%%%%%%%%%%%%%%%%%%%%%%%%%%%%%%%%%%%%%%%%%%%

\section{Data reductions}
\subsection{Calibrations}
All calibrations and data reductions were performed using standard routines in the {\it Common Astronomy Software Applications} package, CASA\footnote{{\tt http://casa.nrao.edu}}, using versions 4.2 and earlier. Paper III describes in detail the process we have been following for calibrations and reductions. A summary follows here: 

In order to reduce ringing across the band from strong RFI, it was necessary to Hanning smooth our data. Antenna-based delay calibrations have to be applied prior to this smoothing, which was done in an initial stage of the calibrations. 

The absolute flux density scale of the data was determined by observing one of two standard primary calibrators; 3C286 or 3C48 (the latter for three of our early right ascension galaxies, NGC~891, NGC~660 and part of the NGC~2613 observations), and applying the Perley-Butler 2010 flux density scale. 
We used the primary calibrator to determine the bandpass corrections as well. See Table~\ref{tab:observations} for a list of the calibrators which were used for each galaxy in the sample.  

\subsubsection{Polarization calibration}
For the polarization calibrations, the primary calibrator was used as a polarization angle calibrator. One of the NRAO standard zero polarization calibrators was used to calibrate instrumental polarization leakage. 
We corrected for cross-hand delays, leakage terms and polarization position angles. Where in some cases the polarization leakage calibrator could not be used (due to loss of scan during observation or being more heavily flagged than the other calibrators), we instead utilized the secondary calibrator where the range in parallactic angle was sufficient, i.e. more than 60 degrees. This was for example done for the galaxies in a C-band scheduling block (ID: 4809749), for which the polarization leakage calibrator scan was lost. 
Q and U fluxes were calculated in the standard way -- see Paper II and references therein. Our tests have shown no difference in the result whether the secondary calibrator or polarization leakage calibrator was used. 

\subsection{Flagging}
For all scans, calibrators and galaxy scans alike, we inspected the uv-data by eye. All 
bad data, whether from RFI or instrumental effects, were manually flagged. A number of tests were carried out using automatic CASA flagging routines as well as the CASA pipeline; however, we found significantly improved results by continuing with manual flagging. 

Generally, fewer antennas could be used for our L-band observations than for C-band, since not all VLA antennas had yet been equipped with L-band receivers at the time of observation. Additionally, there is typically more RFI at L-band. 

Calibrations and flagging were carried out iteratively, until a well-calibrated data set could be obtained. The final calibrated galaxy scans were then separated from the calibrator scans, in order to be imaged in Stokes I, Q and U. 

\subsection{Imaging}
\label{sec:imaging}
%referee wants this much shorter, this and the next 3.3.1 (weightings)
We have produced Stokes I, Q and U images, as well as spectral index maps ($\alpha$, where $I_\nu \propto \nu^\alpha$) and uncertainty maps, $\Delta \alpha$ (see Sect.~\ref{sec:spectral_indices}) from the calibrated data. 

The {\it clean} algorithm in CASA, which we use for our wide-field synthesis deconvolution, allows for multi-scale multi-frequency synthesis, {\it ms-mfs} \citep[see][for a full description]{RauCornwell2011}. % The ms-mfs algorithm enables simultaneous spatial and spectral fitting of wide-band data.  
%While previous mfs deconvolution algorithms used zero-scale (point source) flux components to model the sky emission ('classic clean'), this is not well suited for the extended emission of our galaxies. 
We set the multi-scale feature to look for flux components at a variety of spatial scales, to account for the extended emission of our galaxies. %The user's choice of scales is based on angular size of the source, beam resolution, and estimates on actual sky emission, but can vary with data quality. 
Typically our scales ranged from zero ('classic clean') up to approximately five times the synthesized beam, but sometimes required adjustments. In L-band, and for the uv-tapered weighting in particular, fewer scales were at times necessary in order to remove artifacts for point-like sources; for example a classic clean was set for NGC~2683, NGC~2992, NGC~4157 and NGC~4845. 

Additionally, the {\it clean} algorithm enabled simultaneous fitting of a spectral index across the band width ('in-band spectral index') via a simple power law.  In order for this to take place, we set the number of Taylor coefficients used to model the sky frequency dependence,{\it nterms}, to 2 for all CHANG-ES galaxies. We defer the details of spectral information to Section~\ref{sec:spectral_indices}, and will describe the spatial fitting, as implemented for the CHANG-ES project, in this section.

%We describe the deconvolution process for Stokes I in Sec.~\ref{sec:stokesIimaging} below. After the best settings had been found, the same settings were used for Stokes Q and U (see Section~\ref{sec:polarizationimaging}). 

\subsubsection{Weightings}
\label{sec:weightings}
For each galaxy, the Briggs robust weighting \citep{Briggs} %with the robust parameter set to zero in CASA's CLEAN algorithm 
was used for both C and L band data. %, in order to reach an optimum balance between sensitivity and resolution.
Additionally, we imaged a uv-tapered version of this weighting, which we introduced in order to emphasize the broadscale structures\footnote{For an example of a different weighting, see Paper II for which a naturally weighted map of NGC~4631 was included.}. We achieved an increase in beam size of approximately 80\% in C-band and 40\% in L-band, by typically applying a 6 k$\lambda$ Gaussian taper in C-band and a 2.5 k$\lambda$ Gaussian taper in L-band to the uv-distribution. Image products of both these weightings are included in Data Release 1 and are referred to as robust 0 and uv-tapered weightings, respectively. 

In C-band, images sufficiently clear of artifacts with uv-tapered weighting could not be achieved for NGC~4438 and NGC~4845 due to particularly strong contaminating sources in the field.  In L-band, the uv-tapered weighting was either challenging to achieve (for the same reason) or unnecessary (the sources were imaged as point sources already at robust 0) for six of our galaxies; NGC~3448, NGC~4192, NGC~4388, NGC~4438, NGC~4845 and UGC~10288, and the attempt was discarded for these.  
Hereafter any images shown will be of robust 0 weighting, unless otherwise specified. 

\subsubsection{Stokes I imaging}
\label{sec:stokesIimaging}
Each imaging run was cleaned down to a flux density level of 2.5-3 $\sigma$. With few exceptions, we imaged the entire field (i.e. without specifying regions), since a lower rms could be obtained with this strategy. Additionally, the vast majority of our data do not suffer from significant artefact patterns with spurious sources outside the source region, which could jeopardize this strategy.
When needed, one to two self calibrations were performed in order to deal with remaining artifacts. 

Our self calibration strategy was based on iteratively refining the calibration table, by self calibrating at successively deeper thresholds and thus improving the input model, which in turn acts on the original visibilities. At each step, care was taken to a) check the model to be used for the self calibration that it did not include any artifacts and b) check that the peak intensity of the source would not decline (such a scenario was aided by opting for a phase-only self calibration at the first iteration, or by  applying the self calibration at a shallower threshold). 

In some cases, our cleaning strategy had to be adapted to the circumstances of the data (the quality of a dataset, strength of emission, contamination in the field, etc). This particularly applies to data with high dynamic ranges (HDR), due to galaxies with strong, often point-like centre sources or strong field sources (for example NGC~660, NGC~4594, Virgo galaxies). Our strategies to deal with these situations include one or more consecutive runs of self calibrations, specifying regions, attempts at peeling for strong field galaxies (see \ref{sec:peeling}) and clean parameter adjustments (including multi-scale adjustments). Generally, the best one can achieve in terms of HDR, is a signal over noise of the order of 10000:1 (S. Bhatnagar, priv.comm.). Our dynamic ranges are listed in column 8 of Tables~\ref{tab:imagingparametersC} and \ref{tab:imagingparametersL}. Additionally, in cases of strong centre sources, choosing a smaller cell size will characterize the steep beam better and render a cleaner result (e.g. NGC~4594 and NGC~4845, both in C-band). 

\subsubsection{Polarization imaging}
\label{sec:polarizationimaging}
Stokes Q and U images were produced similarly to I. These maps were used to create polarization intensity and polarization angle maps. The polarization intensity maps were corrected for bias caused by the noise level\footnote{Note that the maps shown in panels (e) and (f) of the images in the appendix are the noise biased maps. Peak polarization values have however been measured from the corrected maps.} and the polarization angle maps are cut off at the 3$\sigma$ level. We note that if the fraction of polarization intensity over Stokes I intensity is less than 0.5\%, the polarization is not believable since the calibration cannot be guaranteed below that level (S. Myers, priv. comm.). 

A uniform Faraday screen resulting from the ionosphere is corrected for in the standard polarization data reduction. However, if there is differential Faraday rotation, for example between the location of the primary calibrator and the source, then this is not corrected for\footnote{The software is not yet available in CASA.}. This effect is negligible at C-band. There could, however, be minor effects at L-band, which will be investigated in a subsequent polarization paper.
 
Our vector maps (panels (d), (e) and (f) of the figures in the appendix) are also 'apparent B-vectors' which simply are E-vectors rotated by 90 degrees, because internal Faraday rotation in the galaxy has not been corrected for. 

\subsubsection{Primary beam corrections}
\label{sec:pbcorrection}
Wideband primary beam corrections were carried out with the CASA task {\it widebandpbcor} after cleaning. This task accesses information about the beam from the visibility data, and calculates a known NRAO-supplied model of the primary beam (PB) at each frequency channel in each band. A PB cube is formed, and Taylor terms are found in the same way as was done for the data set when imaging. Each Taylor coefficient image (two in our case, see Sect.~\ref{sec:imaging} and \ref{sec:spixmaps}) is then corrected. 

This task also corrects the spectral index image above a given threshold (we use 5$\sigma$), as well as corrects a map of the formal error of the spectral index (see Sec.~\ref{sec:spixcorrections}). Subsequently, we also applied primary beam corrections to the polarization intensity maps, by dividing the image with the beam created in the {\it widebandpbcor} task. 

Note that in this paper, any displayed images are not PB corrected, but any measurements are made from the PB-corrected versions of the maps, unless otherwise indicated. For total intensity images, use of a model for the PB rather than an observation-specific PB introduces negligible error over the scales of our galaxies. However, the effect on the spectral index maps must be investigated in more detail, which we do in Sec.~\ref{sec:spixcorrections}. 

\subsubsection{Mosaicking two pointings}
\label{sec:twopointingimaging}
For the eight large galaxies which were observed in two pointings at C-band (see Table~\ref{tab:twopointings}), each individual pointing was calibrated and imaged separately. Although it would be preferrable to mosaic the pointings onto the same grid while still in the uv-plane, this feature is not yet implemented in CASA at the time, for the {\it clean} settings that we required (i.e. {\it nterms = 2}). Instead, we combined the images from the two pointings, forming a weighted (by the primary beam) average in the overlapping region. The images were combined in the overlapping regions out to 10\% of the primary beam.
This rather inclusive choice of cut-off worked well for the D configuration data, but will likely be adjusted to 50\% for the B and C configuration data in the future.

This strategy was applied to both the non-PB-corrected images, as well as the two primary beam pointings. The combined Stokes I image was divided by the combined primary beam in order to do the PB correction. 

The two-pointing spectral index maps, which had already been individually PB-corrected, were similarly combined, as well as their accompanying error maps. 

We assessed whether the two pointings had differences within the overlap regions, such as in noise levels. In such cases additional weightings, based on the noise level, would have to be considered. This was however not the case for any of the CHANG-ES galaxies, where separate pointings generally had been observed in turns in the same scheduling blocks, or in similar conditions. 

\subsubsection{Peeling of strong field sources}
\label{sec:peeling}
For a couple of cases, for example NGC~660 and NGC~4388, bright field sources interfere strongly in the cleaning process, causing a less than ideal imaging result fraught with severe artifacts (L-band only).
An attempt was made at {\it peeling} these interfering sources in those examples, that is, to remove or decrease the intensity of the interfering source so that its impact is lessened in the final deconvolution. The process includes self calibrations with the interfering source centred, and subtracting the source from the uv-data using the CASA task {\it uvsub}, in several steps. For more information on peeling, see \citet{pizzoanddebryn2009}, as well as the PhD thesis by \citet{Adebahr2013}.
Table \ref{tab:peeling} lists the sources we peeled from NGC~660 and NGC~4438. 

%1. clean full image
%2. self cal full image (initial selfcal
%3  clean with box over evil source only 
%4  Use the model from that clean in original image
%4.5. make a selfcal table with the model from teh evil source used in the original
%5. subtract the model from the original using uvsub
%6. Create a new msfile from which the field will be removed
%7. invert the selfcal table from 4.5. 
%8. apply it => removes the field
%9. make an image with the evil source centered (make use of the phasecenter parameter) and selfcalibrate.
%10 Now selfcal with the evil source centred. 
%11 invert this table and apply
%12 now subtract this (evil source) from the original file
%13 and clean... 

%%%%%%%%%%%%%%%%%%%%%%%%%%%%%%%%%%%%%%%%%%%%%%%%%%%%%%%%%%%%%%%%%%%%%%%%%%%%%

\subsection{In-band Spectral Indices}
\label{sec:spectral_indices}

\subsubsection{Formation of Spectral Index Maps}%Judith
\label{sec:spixmaps}
As indicated in Sect.~\ref{sec:imaging}, the ms-mfs algorithm allows for spectral fitting across the wide bands that have been used in the CHANG-ES survey.  Thus, a single observation over a single wave-band permits the determination of spectral indices, $\alpha$, as a function of position 
 within the galaxy. As described in \cite{RauCornwell2011} (see also Paper II), a specific intensity, $I_\nu$, at some frequency, $\nu$, within the band, can be fit with a function of the form 
$I_\nu\,=\,I_{\nu_0}\,\nu^{\alpha\,+\beta\,log\left(\nu/\nu_0\right)}$, where $\beta$ is a curvature term and $\nu_0$ is a reference frequency (our central frequency, $\nu_0$, given in Tables~\ref{tab:imagingparametersC} and \ref{tab:imagingparametersL}).  As implemented, such a fitting function is expanded in a Taylor series about $\nu_0$, such that the first Taylor term (the map, $TT0$) corresponds to a map of specific intensities at $\nu_0$ and the second Taylor term (the map $TT1$) contains information on $\alpha$ such that $\alpha\,=\,TT1/TT0$. The third Taylor term ($TT2$) allows for the recovery of the spectral curvature, $\beta$, where $\beta\,=\,TT2/TT0\,-\,\alpha(\alpha\,-\,1)/2$.

In principle, any number of Taylor terms could enter into such an expansion with increasing numbers of terms improving the spectral fit \citep[see][for examples]{RauCornwell2011}. In practice, however, the effective number of terms is limited by the signal-to-noise (S/N) ratio and calibration precision of the data.  For the CHANG-ES galaxies, tests that we have run gave poorer results with 3 terms than with 2; that is, the rms noise in the images is higher for such fits, when cleaning to the same threshold. 

The curvature maps also show large and sometimes unrealistic variations, point-to-point. While globally averaged values of $\beta$ can still be useful (see Paper II for an example), we have adopted two Taylor terms for the CHANG-ES project as indicated earlier.  A two-term Taylor fit then takes the simple form 
$I_\nu\,=\, I_{\nu_0}\,+\,I_{\nu_0}\,\alpha\,\left(({\nu-\nu_0})/{\nu_0}\right)$. A map of random errors describing the accuracy of the fit, $\Delta\,\alpha$ is also formed\footnote{See Eqn. 39 of \citet{RauCornwell2011}.}.

The beauty of the spectral fitting is that it is carried out by {\it flux component} 
which means that a uniform resolution is achieved over the band, with a restoring beam applied to the TT0 and TT1 maps at the end of the fitting process.  Also, the S/N of the entire band applies to each flux component.  If one were to form a spectral index in the classical way within a band, one would need to break up the band into its various channels (or groups of channels), smooth each channel to the poorest resolution of the lowest band frequency, and accept the very high noise of these channels as a cut-off when forming $\alpha$ maps.    

Clearly, the method, as implemented in CASA, is far superior for any individual band\footnote{For an example showing a result when more than one band is used, see Paper III.}.  Nevertheless, there are limitations, as we outline below.

\subsubsection{Post-Imaging Corrections to the Spectral Indices}%Judith
\label{sec:spixcorrections}
Because the primary beam (PB) varies with frequency, it imposes its own spectral index, $\alpha_{PB}$, onto the $\alpha$ maps.  This effect is significant; for example, at the 50\% level at L-band, $\alpha_{PB}\,=\, -1.6$. We use the CASA task {\it widebandpbcor}, described in Section~\ref{sec:pbcorrection}, to correct for $\alpha_{PB}$, and at this point, we cut off the $\alpha$ maps at 5$\sigma$, where $\sigma$ is the rms of the image map (i.e. the TT0, or $I_{\nu_0}$ map).

Once PB-corrected $\alpha$ maps are formed, they can show large variations around their peripheries and the corresponding error maps, $\Delta\,\alpha$, show correspondingly large errors.  In addition to the 5$\sigma$ cut originally applied, we also cut off all $\alpha$ maps wherever the corresponding error map exceeds a value of 1.0. The choice of this cutoff is arbitrary; however, various tests showed that more aggressive cuts (e.g. where $\Delta\,\alpha\,>\,0.75$) resulted in undesirable `holes' in some $\alpha$ maps. Note that this step does not change the PB correction of the $\alpha$ maps. It simply discards data points that are known to have large errors. 
%Moreover, this decision made negligible difference in our calculation of averaged weighted alpha values (see Sect.~\ref{alpha_weighted}).

Both $\alpha$ and $\Delta\,\alpha$ maps are computed for each map pixel (cell).  However, these pixels are not independent.  For example, at L-band, there may typically be 64 cells/beam and at C-band, 50 cells/beam (both uv-tapered). Consequently, the `per pixel' errors are much greater than the `per beam' errors and the $\alpha$ maps, as produced, show variations pixel-to-pixel which are larger than beam-averaged variations would be.  In addition, maps of $\Delta\,\alpha$ show artifacts, i.e. regions of low $\Delta\,\alpha$ that have a `thread-like' appearance through the map.  These artifacts depend on the choice of multi-scales that are used during imaging and shift position if the map is remade with different scales. If the adopted scales are changed, these 'threads' also shift position.

To ameliorate the above issues, we have `smoothed/ averaged' our maps of $\alpha$ and $\Delta\,\alpha$ over the size of a beam.  This is accomplished by artificially inserting into the map header a Gaussian beam size whose area (Gaussian-weighted) is equivalent to the pixel area and then convolving the map to the correct clean beam size\footnote{The CASA task {\it imsmooth} is used.}. The result does {\it not} change the spatial resolution of the images but recognizes the non-independence of the pixels and strongly minimizes pixel-to-pixel variations. For example, the rms of an $\alpha$ map could decrease by as much as a factor of 2 after such a convolution.  These are the final maps that are shown in the appendix, panels h), i), k) and l).  

In summary, our final spectral index maps and all calculations performed on them apply to $\alpha$ after 
\begin{enumerate}
\item PB-correction, 
\item a 5$\sigma$ cutoff, 
\item a cutoff wherever $\Delta\,\alpha\,> 1.0$, and 
\item convolving with a Gaussian to smooth over pixel-to-pixel variations within any clean beam.
\end{enumerate}

\subsubsection{Uncertainties in Spectral Index Maps}
\label{sec:alpha_errors}
As indicated above, $\alpha$ is determined via fitting flux components over the band during the imaging process.
Fitting errors should, for the most part, be accounted for by the $\Delta\,\alpha$ maps, but those maps consider random errors only.
Therefore, a variety of tests have been carried out on the in-band spectral index maps to help us understand realistic uncertainties, and we outline them here.
Largely qualitative descriptions are provided, with a quantitative summary at the end.
In the following, we refer to the final $\alpha$ maps, as described above, unless otherwise indicated.

\noindent{\it Edge-effects:}
 
In spite of the cutoffs and smoothing outlined in the previous section, values in the spectral index maps (as well as their errors) tend to become extreme at the edges (for example, see the L-band panels (i) and (l) of NGC~3735 in the appendix). 
%We have left these edge-effects as is, but note that they make a negligible difference to the weighted average $\alpha$ values (see Section.~\ref{alpha_weighted}) provided in Table~\ref{tab:weightedalpha}. 

\noindent{\it Primary Beam Correction:}

The NRAO-supplied PB model cannot take into account variations in the PB that are specific to any given observation.  The true PB will change with flexure in the antennas and therefore with position on the sky.  It will also rotate on the sky as the source is tracked and therefore any non-axisymmetry in the PB will also introduce error\footnote{Recently, it has become possible to correct for narrow-band time-varying PB effects during the imaging process using the `A-Projection' algorithm \citep{Bhat08, Bhat13}. However, an algorithm applicable to CHANG-ES data (i.e. for wide bands) is not yet available or practical.}.  As a result, PB correction errors, which increase with distance from the pointing centre, must be understood.  \citet{Bhat13} have shown that errors in the $\alpha$ maps are negligible out to the half-power point but increase significantly beyond that distance (see their Fig.~4, right).  For L-band, our largest $\alpha$ map (NGC~4565) extends to a radius of 6.8 arcmin which is well within the HPBW of the L-band PB (radius of 15 arcmin, or FWHM of 30 arcmin). At C-band, however, the PB radius is only 3.75 arcmin and consequently $\alpha$ maps with radii larger than this will show such errors in their outer regions.  

We have investigated such potential PB errors in two ways:  

The first is to examine $\alpha$ maps using cases in which the observations of a given galaxy were carried out in two different observing sessions, such as NGC~660 (see Table~\ref{tab:observations}).  As such, it is possible to form two different $\alpha$ maps for the same galaxy and same sky pointing but observed on different days.  This has allowed us to compare the resulting $\alpha$ maps and to compare the variations between those days with the random errors as given in the $\Delta\,\alpha$ maps.  We tested both low and high dynamic range cases and compared the two results quantitatively.

The second is to examine our C-band observations of large galaxies where two offset pointings were observed (see Sec.~\ref{sec:twopointings}). 
A comparison of $\alpha$ made for the individual pointings allowed us to investigate $\alpha$ errors at larger distances from the PB centre.

%%%%%%%%%%%%%%%%%%%%%%%% Added Philips text here!
As an example, Fig.~\ref{fig:Philipfig} (panel a) shows a map of the absolute pixel-by-pixel difference between the PB-corrected spectral index measurements in the two C-band pointings of NGC 891. An increase of the differences with distance from the map centre is clearly apparent. 
We formed the spectral index error (shown in panel b) by computing the PB-weighted deviation from the average  spectral index of the two pointings\footnote{
$\sqrt{(w_1(\alpha_1 - \alpha_{avg})^2+w_2(\alpha_2-\alpha_{avg})^2)/(w_1+w_2)}$, where $w_1=beam_1/beam_2$ and $w_2=beam_2/beam_1$}. 

This error map is for the most part comparable to the error calculated by the ms-mfs algorithm (panel c)), as the map of the ratio of these two errors (panel d)) illustrates. A major exception to this is the disk of the galaxy, but only in those parts that are located outside the 70\% PB level (i.e. where the PB gain is less than 0.7) of either pointing. The average error ratio in these narrow regions around the mid-plane is $\sim$5 (with individual pixel values up to $\sim$15), i.e. here the error originating from the pointing differences is on average $\sim$400\% higher than the error resulting from the ms-mfs spectral fitting (for comparison, the average error ratio of NGC~4565 is 3.3). Such high ratios are primarily a consequence of the small ms-mfs-based errors in regions of high signal-to-noise, yet the increase of the pointing differences with distance from the pointing centres is significant. In particular, beyond the half-power point the mid-plane errors in panel b) of Fig.~\ref{fig:Philipfig} increase up to $\sim$0.5, whereas in panel c) they remain below 0.1 throughout the disk, as shown by the displayed contours. While the errors in panel b) are in rough agreement with \citet{Bhat13} in the sense that they do not exceed 0.1 out to approximately the half-power point (not considering the above-mentioned edge effects), the error ratio map suggests that the simple PB model used by the widebandpbcor task already shows significant inaccuracies at the 70\% level.

\noindent{\it Resolved out structures:}

Although the method of calculating the spectral index ensures that the resolution is common across the band, if there are structures that are resolved out at one end of the band compared to the other, this is not accounted for in the spectral index maps. For example, there may be structures resolved out at the high frequency end of the band, but not at the low frequency end, which would steepen the spectral index. %This can not be avoided unless the zero spacing flux is added in, a step which we hope to accomplish with Green Bank Telescope data in the future. %This should only matter for the larger galaxies and the importance of the effect can be estimated by comparing our flux densities with previous single dish values. 

\noindent{\it Comparison with Classical Spectral Index:} 

A comparison was made between the in-band spectral index and a classically formed spectral index from a given band.  The classically formed $\alpha$ maps were
made from the spectral index maps after the PB-correction was applied but not after applying further processing (i.e. PB-correction and a 5$\sigma$ cutoff were applied, but not a cutoff based on the error map, nor smoothing to the size of the beam, as described in Section~\ref{sec:spixcorrections}); they were formed by imaging the lowest end of the band and the highest end of the band, smoothing to a common spatial resolution
and then combining in the classical way.
The results were in agreement within errors.  

\noindent{\it Two-pointing PB cutoffs:}
For `two-pointing' galaxies (large galaxies at C-band) the maps, as indicated in Section~\ref{sec:twopointingimaging}, 
were combined over regions in which the value of the primary beam exceeded 0.1 (10\% of the peak). When mosaicking is carried out, each point is weighted by the PB such that points that are farther from the beam centre are weighted down and therefore the difference in total intensity maps, whether one cuts off points where the PB is $>$ 10\% or where the PB is $>$ 50\%, is entirely negligible.
However, since $\alpha$ maps are known to increase quite strongly below the 50\% level, we remade our mosaicked $\alpha$ maps with a 50\% cutoff to compare it to the 10\% cut-off values used in this paper.  For these results, we found only minor differences (for example, if the two maps are subtracted, the rms in the regions between the 50\% and 10\% PB cutoffs is $<$ 0.1).

\vskip 0.2truein
In summary, our final $\alpha$ maps should present realistic results with the following cautions: \\
a) Extreme values around the edges are artifacts and should be ignored.\\
b) Measurements of $\alpha$ should not be quoted `per pixel' but rather averaged over a beam when carrying out evaluations or comparisons. \\
c) The calculated $\Delta\,\alpha$ maps typically {\it underestimate the true errors by $\approx$ 20\%} for galaxies of small angular size\footnote{Estimated from results where galaxies have been observed in two separate observations.}.\\
d) The largest source of uncertainty in $\Delta\,\alpha$ relates to the PB model correction away from the pointing centre, which most strongly affects our largest galaxies in C-band. Spectral index errors increase significantly with distance from the pointing centre, such that beyond the 70\% PB level of either pointing (i.e. at distances greater than 2\arcmin 25\arcsec\ from the respective pointing centre at C-band) the $\Delta\,\alpha$ maps typically underestimate the true errors {\it in the disk} by a factor $\sim$5 (see Fig~\ref{fig:Philipfig}).

%%%%%%%%%%%%%%%%%%%%%%%%%%%%%%%%%%%%%%%%%%%%%%%%%%%%%%%%%%%%%%%%%%%%%%%%%%%%%

\section{Results and data products}
A wide range of data products are created for each galaxy. We have assembled a selection of these for each galaxy, which are presented in the appendix. See the appendix for a detailed listing of each panel. In summary, the 12 panels show: 
\\
(a) and (b): the Stokes I maps for 2 C-band weightings (robust 0 and a uv-tapered weightings), 
\\
(c): Stokes I map for one L-band weighting (robust 0 only), 
\\
(d): an optical image with contours and apparent B-vectors, both from the uv-tapered weighting of C-band 
\\
(e) and (f): polarization intensity maps with apparent B-vectors overlaid for C and L-bands, respectively
\\
(g): a composite image of different weightings/bands (see \ref{sec:jayanneimage}), 
\\
(h), (i), (k) and (l) spectral index maps (h, i) with corresponding error maps (k, l) in C and L-bands respectively. 
\\
(j) a wide view of the L-band field. 
\\

Tables~\ref{tab:imagingparametersC} and \ref{tab:imagingparametersL} list beam sizes, rms noises and dynamic ranges for both bands and weightings. 

\subsection{Rms noise}
\label{sec:rmsnoise}
All rms noise values mentioned throughout this paper have been measured from non-primary beam corrected maps. 
For C-band, the rms noise is measured as an average of the regions throughout the imaged field of view which do not contain detectable background sources.

For L-band, however, there are so many background sources near the galaxy that it is difficult to find sufficiently large regions within which the rms can be consistently measured. Residual cleaning artifacts also tend to be larger at L-band, especially close to the galaxy where the primary beam response is high. Therefore the rms tends to be variable when measured near the source, but declines to more consistent values with distance from galaxy as the primary beam response declines. For consistency, we quote rms values far from the galaxies where there are smaller variations in rms between measurement regions. Rms levels near the galaxy could be up to a factor of two higher than the quoted values of Tables~\ref{tab:imagingparametersC} and \ref{tab:imagingparametersL} but it is apparent from the panel c) of the appendix figures, that our choice of 3$\sigma$ as the lowest contour generally well represents the faintest believable emission.

\subsection{Displaying the halo} 
\label{sec:jayanneimage}
Panel (g) of the figures in the appendix provides an exploratory image intended to help the viewer discover information about individual radio halos. For example, one may like to know if point sources in the disk (anti-)correlate with very extended halo emission.  The CHANG-ES survey is data rich, with two (or more) weightings  provided per band and overlaying contour plots of these for both bands can result in confusing diagrams. Therefore we overlay colourized "transparent" images of the weightings so that the viewer may relatively quickly apprehend the relationship between structures in the two observing bands.  Additionally,  artifacts, confusion structures, and non-random noise are difficult to remove mathematically from radio data. Our visualization approach uses masking in order to mitigate "background" artifacts. 

First the fits data are stretched using the KARMA\footnote{See http://www.atnf.csiro.au/computing/software/karma/} visualization package's {\it kvis} task. On intensity-inverted, logarithmically scaled data, we use the "greyscale 3" option in the pseudo-colour tool for adjustments.  The resulting images are saved in eps format and used as input into the Gnu Image Manipulation Program (GIMP)\footnote{See gimp.org}. This package allows the user to stack images in "layers" that can be combined, as if they are transparent, using a variety of blending mode algorithms. Usually we stack four images, i.e. two weightings of each observed band.  We colourize the C band data (blue for the robust 0 weighted data and green for the uv-tapered weighting) while the L band data are left as greyscale images.  The order in which the images are combined, and which algorithms are used,  are described in  Fig.~\ref{fig:jayanneflowchart}. In the resultant qualitative image, often both point sources in the disk and diffuse halo structure are evident simultaneously. 
s
%%%%%%%%%%%%%%%%%%%%%%%%%%%%%%%%%%%%%%%%%%%%%%%%%%%%%%%%%%%%%%%%%%%%%%%%%%%%%

\subsection{Comments on individual galaxies}
\label{sec:galaxydescriptions}
A few galaxies warrant some extra comments, either because of differences in the data reduction procedures, or because the results were interesting or unusual and therefore caught our attention. In this section, we list these galaxies with comments. Note that this is not meant to be a thorough discussion of each galaxy. 

\subsubsection{NGC 660}
The data of polar ring galaxy NGC~660 have the highest dynamic range in both bands, due to a strong central source. Consequently, the resulting images have higher than expected rms values, particularly in L-band. Nevertheless, the achieved signal to noise of the C-band map with robust weighting reached 70000, on the higher end of what best can be expected (see Section~\ref{sec:stokesIimaging}).

In spite of careful cleaning, including peeling\footnote{Only the Stokes I and spectral index images were produced from the peeled data, while polarization images were not, since the off-centre confusing sources are not as problematic in Q and U as for total I.} of the nearest of the two strongly interfering field sources in L-band (see Section~\ref{sec:peeling}), artifacts are still present. This could potentially be affecting the spectral index results, where regions of flat spectral index are crossing the disk. However, these regions could also be an effect of the possible AGN (see \ref{sec:AGNcontamination}). %Additionally, NGC~660 is considered to be a polar ring galaxy, formed in a dramatic collision a few billion years ago \citep[e.g.][]{combesetal1992, vandrieletal1995} and we could also be seeing an indication of the subsequent enhanced star formation in its flat spectral index. 

We note that the polarization intensity is low, with its fraction of polarization over Stokes I intensity just a 10th of the 0.5\% which is considered a believable signal (see Section~\ref{sec:polarizationimaging}).

%SF or AGN would have flatter spix. Young SB has flatter than older (steepens with age). AGN flat even positive. SB can't result in much of a positive spix. 

%From Marita: 
%http://adsabs.harvard.edu/cgi-bin/nph-data_query?bibcode=1992A%26A...259L..65C&db_key=AST&link_type=ABSTRACT&high=54d367414b21095
%http://adsabs.harvard.edu/cgi-bin/nph-data_query?bibcode=1995AJ....109..942V&db_key=AST&link_type=ABSTRACT&high=54d367414b15142
%In the second, on p. 955, it is claimed that the N660 system was formed in a dramatic collision between two equally massive objects a few billion years ago.
%I also saw it is called a LINER. At least there is CO outside the disk of the galaxy and nuclear activity. Both could lead to a flatter spectral index (enhanced star formation after encounter and probably still today??) or jet activity (I don't know whether indications for this are found yet). 

\subsubsection{NGC~3556 and NGC~5775}
\label{sec:n3556}
%uvtap: There are areas above and below the disk with a lower mean value than the other background, an indication of missing spacings
Both of these galaxies are well-known to have significant halos, which we can also see in our data. Moreover, they both show very flat spectral indices at C-band throughout the disk, where the errors in spectral indices are the lowest. Neither data set showed any particular problems and were single pointings. Bearing in mind the summary regarding spectral index in Section~\ref{sec:alpha_errors}, this suggests that thermal emission may be more strongly dominant for these galaxies in their disks. 

For NGC~5775 at L-band, some broadscale polarization features have been found in the field that are most likely from foreground Galactic emission (this is more obvious when a larger field than shown in the appendix is displayed). The Galactic coordinates of this source place it almost directly over the Galactic centre, although at reasonably high Galactic latitude.
%George's text:
Indeed, the Galactic coordinates of NGC 5775 $(l=359.4^{\circ},b=52.4^{\circ})$ indicate that we view it through the northernmost tip of the Fermi bubbles that extend 55 degrees above and below the Galactic centre \citep{Ackermannetal2014}. Substantial polarized emission at 2.3 and 23~GHz has been found to coincide with the Fermi bubbles, including ridge-like filamentary structures crossing through this particular location \citep{Carrettietal2013}. Our polarization images are likely picking up this same extended foreground structure at lower frequency.

\subsubsection{Three Virgo galaxies: NGC~4192, NGC~4388 and NGC~4438 }
These galaxies are highly affected by contamination from strong sources in the field, such as M87, rendering higher than normal rms noise and artifacts hard to clean out. 
This may have an effect on both spectral index results and polarization. 

We note that the C-band spectral index is steep in the inner disk for NGC~4192 (a rather weak source at the low end of our flux density cutoff for the survey), and that the uncertainties are high. 

NGC~4388 was particularly affected by M87, and we made an attempt at peeling the source in L-band, as well as self calibrating on M84 (the other disturbing source in the field), which rendered some improvement. The resulting rms at L-band was too high for us to detect significant polarized flux (panel f). 

Also NGC~4438 is strongly affected by residual side lobes from M87. Despite attempts of self calibration and peeling, the rms could not be brought down to lower than almost three times the expected value.  
L-band observations of NGC~4438 differ from the other galaxies, in that only the upper half of the frequency range (spectral windows 16-31) was used for imaging, with a central frequency of 1.77 GHz. Q and U images were made over the same upper half of the frequency range in order to match the total intensity image, as well as the spectral index image. 

\subsubsection{NGC 4594}
The strong centre of NGC~4594 (the Sombrero galaxy) resulted in residual side lobes which made it difficult to detect the weak east-west disk for the two-pointing data set in C-band. Careful selection of self calibration inputs eventually revealed the disk, and subsequent merging of the two pointings further helped to cancel out artifacts from the centre. 

In L-band, a plume is seen to the north of the core and is roughly in the direction of the jet observed at much higher resolution by \citet{hadaetal2013}. An unusual polarization structure and a flat spectral index associated with the core and towards the north are also observed. A more conservative spectral index cut-off could be beneficial for this data set (for example, a 10$\sigma$ cut-off was adopted for the in-depth study of NGC~4845, see Irwin et al. 2015 (accepted by ApJ).

\subsubsection{NGC 4631}
The C-band spectral index shows an unexpected flattening from the central to the outer disk for this two-pointing data set. This gradual deviation has not been detected in previous observations of this galaxy \citep{HummelDettmar1990}, but is within the errors pointed out in point d) of Sec.~\ref{sec:alpha_errors}.

\subsubsection{NGC 4666}
Despite its small angular size well within the symmetric primary beam, the C-band spectral index displays an asymmetry between the two halves of the disk, such that it is flatter on the south-west side compared to the north-east. This asymmetry corresponds to an asymmetry in the polarization at the same frequency. Interaction with a nearby companion, NGC~4668  (to the south-east) could be a factor, and/or motion through an intergalactic medium. 
No other technical issues (antenna flexure, solar influence) are problematic. 

\subsubsection{NGC 4845}
It is worthwhile to note that this galaxy displays an interesting variation in flux density level and further exploration, being published in CHANG-ES V (Irwin et al., 2015, accepted by ApJ), indicates with little doubt that the source is indeed variable. CHANG-ES V also explores the spectral index and polarization results in more detail. 

\subsubsection{NGC 5084}
Due to the low declination of this galaxy, the C-band observations were divided up into three scheduling blocks in order to a) avoid shadowing and b) increase uv-coverage. The scheduling block in which the first pointing was observed did not have a polarization leakage calibrator scan, and instead the secondary calibrator used for NGC~4192 was used to calibrate polarization leakage (the secondary calibrator of NGC ~5084 did not have a sufficiently large parallactic angle span). Part of the second pointing was strongly affected by shadowing and only 21 antennas were used for these scans. 

The C-band images show emission extending east and west of the point-like centre, as well as southward from the eastern extension -- these are likely real, since their intensity is greater than 10$\sigma$ (and they roughly follow the disk). However the extensions seen above and below the centre may be spurious since they seem to align with weak residual side lobes.
In L-band, we find no apparent polarization from the source.
The spectral index is flat in L-band and flat to slightly positive in C-band (but the positive trend coincides with higher error values). 

\subsubsection{UGC 10288}
UGC~10288 fell within the intensity cutoff of the CHANG-ES sample by piggybacking on the previously unresolved strong background source to the west of the disk. We refer to Paper III for full details on the observations and analysis of this galaxy. 
 
%%%%%%%%%%%%%%%%%%%%%%%%%%%%%%%%%%%%%%%%%%%%%%%%%%%%%%%%%%%%%%%%%%%%%%%%%%%%%

\section{Analysis}
\subsection{Flux densities}
The flux densities listed in Columns 2 and 4 of Table~\ref{tab:fluxdensities} are the average of measurements taken at the two weightings (robust 0 and the robust 0 with a uv-tapering applied). Our measurements avoided disconnected point-like features immediately around the source but did include anything that might be considered an 'extension' (at this point it is unknown whether the extension is indeed related, or not). For example, NGC~3448 shows clearly a point source at C-band, but the same is not visible as a distinct source at L-band. The error (Columns 3 and 6), which is the difference between the two weightings, also encompasses inclusion vs non-inclusion of such point sources. 
In most cases, however, the calibration error (flux density scale accuracy) which is estimated to be about 2\%\footnote{https://science.nrao.edu/facilities/vla/docs/manuals/\\
oss2014a/performance/fdscale} is larger than this error. The error columns in the table cite the {\it larger} of the two uncertainty estimates. 

For some cases, such as NGC~5907, a background source is visible through the disk -  in such cases, the flux density from these sources has been removed from the total flux density measurement. 

Our new flux density values are the best so far at these two frequencies (cf. Paper I).

%Comments from Judith: Since the regions around Lband and Cband are not exactly the same, it is perhaps not wise to use these values for global spectral indices (though in most cases they should be close).  In addition, we are likely missing some flux for the larger galaxies at Cband.
%%%%%%%%%%%%%%%%%%%%%%%%%%%%%%%%%%%%%%%%%%%%%%%%%%%%%%%%%%%%%%%%%%%%%%%%%%%%%

\subsection{Star formation rates} 
Star formation rates and "surface densities" (see below) are based largely on WISE \citep{Wrightetal2010} 22-\micron\ images with resolution enhanced over the standard survey products to a final value of 12.4\arcsec\ via the WERGA (WISE Enhanced Galaxy Resolution Atlas) process \citep{Jarrettetal2012, Jarrettetal2013}.  WERGA images in all four WISE bands were kindly provided by T. Jarrett.  
Foreground stars were removed through PSF subtraction in the WERGA pipeline.  Backgrounds/foregrounds are relatively uniform, although one galaxy showed a significant jump in background level at 22 \micron\ close to the galaxy but not crossing it.  A constant background was therefore subtracted.  Flux densities were measured with apertures large enough to include all emission from each galaxy and converted to Jy using conversions given in the WISE Explanatory Supplement  \citep{Cutrietal2011}.  Flux densities from any background sources were subtracted.  Three corrections to the flux densities, as described by \citet{Jarrettetal2013} are applied; an aperture correction for extended sources, a small colour correction appropriate for dusty, star-forming galaxies, and a calibration correction at 22 \micron\ appropriate for spiral galaxies.  Formal uncertainties on the flux densities are small, and the final uncertainties include a 1.5\% flux calibration uncertainty \citep{Jarrettetal2013}, although by far the biggest source of uncertainty in the SFR is the galaxy distance.  No corrections were made for extinction.

Spitzer MIPS 24-\micron\ images \citep{Riekeetal2004}, at a resolution of 5.9\arcsec\ and with a range of sensitivities, were found in the NASA/IPAC Infrared Science Archive for 16 of our galaxies, and will mostly be used in future work, but are employed for one purpose below -- hence, their reduction is also described here.  The "pbcd" mosaics produced by the standard MIPS pipeline reduction were found to have low-level "jailbars" in most cases.  We therefore reformed the mosaics from the individual "bcd" images via "self-calibration" (dividing each bcd image by the median of all images) with the MOPEX software. This procedure removed the jailbars very well. No other image artifacts were noted.  Multiple resulting mosaics were added to produce final images. Background gradients were then subtracted with the IRAF program {\it background}.

\citet{Jarrettetal2013} have shown that WISE 22-\micron\ fluxes ($\nu F_\nu$) correlate extremely well with Spitzer 24-\micron\ fluxes for their sample of 17 nearby galaxies, allowing star formation rate calibrations developed for the latter by \citet{Riekeetal2009} to be applied to the former.  
We therefore use the relation between 22-\micron\ $\nu L_{\nu}$ and SFR given by \citet{Jarrettetal2013} to calculate SFRs, except for NGC 4388 where the \citet{Riekeetal2009} relation is used.  

Fig.~\ref{fig:spitzerwise} shows the same 22 vs 24\micron\ flux relation for the CHANG-ES galaxies with MIPS images, and its obvious correlation further encourages this approach. 

To measure 22-\micron\ galaxy isophotal diameters, we formed a major axis profile averaged over a 13\arcsec\ minor axis extent, and measured the diameter where the flux had dropped to three times the typical uncertainty level in the images.  This flux level is 15.7 $\mu$Jy/arcsec$^2$.  A measure of the star formation rate surface density was then calculated by assuming disk axisymmetry and dividing the SFR by the isophotal disk area.
Flux densities, angular diameters, SFRs and SFR surface densities are listed in Table~\ref{tab:sfr_etc}.

\subsubsection{AGN contamination}
\label{sec:AGNcontamination}
About ten galaxies have bright nuclei at 22 \micron, raising the possibility of significant AGN contamination. To give an idea of the potential contamination, we refer to the study of the AGN contribution to mid-IR continuum emission based on observations of the nuclei of nearby AGN by \citet{Tommasinetal2010}. They find lower limits of 45-73\% for the AGN contribution to the 19 \micron\ continuum, depending on the type of AGN. For cases where we believe an AGN to be present, we crudely correct our SFRs by assuming that 100\% of the nuclear 22\micron\ emission arises from the AGN. 
A literature search suggests that all ten candidates are starburst nuclei except for two, NGC~3735 and NGC~4388, with perhaps the most compelling evidence coming from the IR line ratio study of nuclei by \citet{PereiraSantaellaetal2010}.  

We explored this issue further by making a rough determination of  the nuclear 3.4-4.6 \micron\ colour from the WISE WERGA images, a measure which distinguishes well between AGN and starbursts \citep[][Fig. 26, and references therein]{Jarrettetal2011}. We estimated the colour of the central source in two different ways, both of which include simplifying approximations. The resolutions are 5.9\arcsec\ and 6.5\arcsec\ at 3.4 and 4.6 \micron\, respectively.  First, we fit a 2-D Gaussian with a constant background to the central peak in each band. We tried this with two different box sizes: 9\arcsec\ x 9\arcsec\ and 12\arcsec\ x12\arcsec.  We then turned the flux density ratio of the Gaussians into a colour. The second method was to Gaussian-convolve the 3.4 \micron\ images to the 4.6 \micron\ resolution, make a colour map, and examine the colour in a central box of size 6\arcsec\ x 6\arcsec. The reason these methods are approximate is first that the PSFs have significant Airy disks, so some of the flux density is not in the central peak and not described by a Gaussian, and second that a constant background is not necessarily the best assumption in the first method since the galaxy has structure as it cuts through the fitting box. However, the results are reasonably consistent from the two methods.

Although Fig.~26 of \citet{Jarrettetal2011} represents a range of redshifts, our resulting colours clearly fall into either the spiral/starburst or AGN regions, and the classifications are generally consistent with the evidence from the literature, with two exceptions.  

The first exception is NGC~3735, which has a starburst-like nuclear colour of about 0.0-0.3 mag (the range is from the two different methods and shows the greatest discrepancy for any of the ten galaxies), whereas MIR line ratios from \citet{PereiraSantaellaetal2010} clearly indicate an AGN. In this case, we tend to favour the MIR over the NIR evidence because of the possibility of extinction hiding the NIR emission from the AGN in an edge-on geometry. If it is an AGN, we can only estimate the SFR by measuring the disk flux density outside a radius of 13\arcsec. This reduces the SFR to 1.1$M_\sun$/yr, although strictly this is a lower limit. 

The second exception is NGC~660, where a colour of 1.2-1.3 clearly indicates an AGN. If indeed an AGN, the nuclear flux density can be crudely subtracted via a Gaussian fit to the central peak in the 22\micron\ image, leaving a lower limit SFR of 0.61-0.81 $M_\sun$/yr, depending on the fit parameters. %This estimate is a lower limit because some of the nuclear flux may still be due to star formation. 
Yet the literature \citep[e.g.][]{PereiraSantaellaetal2010, BernardSalasetal2009}, indicates a starburst, as well as a LINER galaxy.
Thus, we will continue assuming NGC~660 to be a starburst rather than AGN galaxy, and retain the originally derived values in Table~\ref{tab:sfr_etc}. 

The nucleus of NGC~4388 does appear to be AGN-dominated \citep[see also e.g.][]{Falckeetal1998}, with a colour of 1.1 mag. The 22-\micron\ flux density is dominated by this source, and the faint disk emission is not well resolved from it. Hence, to estimate the SFR, we turn to the Spitzer 24-\micron\ image, where the disk is better resolved.  However, in this case the nuclear source so overwhelms the disk that the uncertainty in the flux density of a Gaussian fit precludes an estimate of a disk SFR as was done for NGC~660. At best, we can crudely estimate a disk flux density outside a radius of 7\arcsec\ from the nucleus.  Hence, the SFR of 0.07 $M_\sun$/yr is a lower limit.

\subsection{Overview of the CHANG-ES sample ordered by SFR surface density} %Judith's
\label{megamaps} 
It has long been suspected that the extent of radio halos is related to SFR or SFR per unit area (SFR surface density) \citep[e.g.][]{Dahlemetal2006}. As a preliminary `quick-look' to see whether such ordering applies to the CHANG-ES sample, we have formed two maps displayed in Figures~\ref{fig:megamap1b} and ~\ref{fig:megamap3}. These maps have been formed from the L-band primary-beam-corrected images\footnote{The scaling described in this section was carried out using the Astronomical Image Processing System (AIPS) of NRAO.}. They have then been ordered by SFR surface density 
as described in the previous section, the latter quantity importantly being independent of distance.

It is important to keep in mind in this section, that in spite of the well-defined criteria used in the previous section, some starburst-classified galaxies may still have, in addition, hidden AGN contamination, and future examination of the CHANG-ES sample including higher resolution data sets should help to clarify this issue.

The first map  (Fig.~\ref{fig:megamap1b}) is an attempt to bring all galaxies to a common distance of 10 Mpc so that their 1.6 GHz sizes can be compared.  Each image was first rotated according to the Ks position angle provided in NED and then minor adjustments were made of order $<\,|4^\circ|$ to produce better horizontal alignment. The galaxies were then scaled in size by the ratio, (D/(10 Mpc))$^2$. Appropriate adjustments were made to ensure that the brightness (i.e. specific intensity) did not change during this process.
The displayed beams (lower right hand corner of each panel) were scaled in the same way and, since the resolutions were all similar (though not identical) for the L-band  data, a glance at these beams reveals the distances and linear spatial resolutions of the galaxies; the more distant galaxies have large beams after scaling, whereas the closer galaxies have small beams.

Fig.~\ref{fig:megamap1b} reveals a wide range of physical sizes for the radio continuum emission in the various galaxies, both along the disk and into the halo. Some of the galaxies show a strong dominant central radio core (e.g. NGC~4594), others show a fairly large vertical radio extent (e.g. NGC~4666);  some appear rather long and thin (e.g. NGC~4565), others show peculiar structure (e.g. NGC~4438 which has a radio lobe) and yet others are clearly influenced by nearby companions (e.g. NGC~4302).

The physical resolution, of course, varies for each galaxy and a detailed analysis must await a future paper.  Nevertheless, it is fairly obvious that the first map (Fig.~\ref{fig:megamap1b}) is more of a comparison of radio continuum sizes than a revelation about correlations with SFR per unit area. We also note that there is sometimes an uncomfortable range of distances in the literature for some of the galaxies.

The second map (Fig.~\ref{fig:megamap3}) is again ordered by SFR surface density.  However, rather than scaling by distance, the galaxies are scaled by the 22 \micron\ WISE WERGA major axis sizes (Table~\ref{tab:sfr_etc}). The galaxy which has the largest angular size on the sky is NGC~4244 and the smallest is NGC~5084 (11.5\arcmin\ and 0.73\arcmin\ respectively).  All galaxies were therefore scaled up to match the angular size of NGC~4244, again ensuring that the brightness was not changed and that the beams were also scaled. Such a map now essentially corrects for the various physical sizes of the galaxies (at 22 \micron).

In this map, the radio emission still shows some size variation because the radio emission embedded within the galaxy might be compact or extended and, again, the spatial resolution varies.  Nevertheless, a trend is now apparent, with galaxies at the top left (high SFR surface density) appearing more `boxy', on average, than galaxies at the bottom right (low SFR surface density).  For example, NGC~4244, which has the lowest SFR/unit area is clearly a very `flat' galaxy with no radio halo evident, consistent with the \textsc{H\,i} distribution observed with a deep observation by \citep{Zschaechneretal2011}.  On the other hand, NGC~4666 which is unlikely to be contaminated by an AGN and has good spatial resolution, has an obvious halo.

Explorations of correlations with other parameters awaits a future paper.

\section{Median edge-on galaxy and its halo}
The fact that we now have all maps scaled to the same 22\micron\ diameter in Fig.~\ref{fig:megamap3} allows us to form the median edge-on galaxy at L-band. That is, we can explore what the typical edge-on spiral galaxy looks like in the radio continuum. We have done this for 30 of our 35 galaxies, excluding NGC~660 (too distorted since it is a merger), NGC~4438 (presence of plume/bubble), NGC~4594 and NGC~5084 (large beam), and UGC~10288 (emission dominated by a background AGN). 
 
We formed the median of the 30 galaxies after having converted the brightness to units of Jy/pixel. The result is shown in Fig.~\ref{fig:mediangalaxy} and takes advantage of the collective sensitivity of all 30 galaxies. %In addition, we checked the contribution of the disk emission to the apparent vertical radio extent by examining the lower inclination galaxies in the sample, finding that the disk made a negligible contribution. 
In the image, we also plot a sample 22\micron\ contour (red) that corresponds to the scaling of the radio data. 

Fig.~\ref{fig:mediangalaxy} shows the spectacular radio extent of the typical spiral galaxy (note that the average beam is much smaller than the displayed radio extent), predicted half a century ago by \citet{GinzburgSyrovatsky1961} (see their Fig.~1, p. 18). The galaxy has the appearance of a slightly flattened ellipsoid and reveals that cosmic rays and magnetic fields not only permeate the galaxy disk itself, but extend far above and below the disk, as has been discussed by e.g. \citet{HaverkornHeesen2012, Krause2009}. %references added
We note that the red ellipse is not affected by the larger beams of the radio images and should not be used, at this point, to make conclusions about the radial distribution of the emission compared to the 22\micron\ emission. On the other hand, the conclusion of a broadscale halo is robust; we have checked the contribution of the disk emission to the apparent vertical radio extent by examining the lower inclination galaxies in the sample, finding that the disk made a negligible contribution. If galaxies that have lower inclination are removed, the halo remains, suggesting that the vertical distribution is not a result of a projected disk. In addition, the halo extends beyond what would be expected from beam smoothing. 

The galaxy scale heights \citep[see][for method]{Dumkeetal1995} and magnetic field extents will be discussed further in future papers. 

\section{Conclusions}
\begin{enumerate}
\item In this CHANG-ES IV paper, we present all the VLA D-configuration observations and results of the 35 edge-on CHANG-ES galaxies in two frequency bands, C and L. These data products (including intensity maps, spectral index maps, polarization maps)  areall be part of our Data Release 1, located at http://www.queensu.ca/changes. Apart from presenting each galaxy in a range of maps for two different weightings of each band, we also investigate deeper the in-band spectral index maps, primary beam corrections, flux densities and star formation rates.  
\item Spectral index maps presented have a) been primary beam corrected, b) cut off below 5$\sigma$, c) cut off wherever the formal error $>$ 1.0, and been convolved with a Gaussian to smooth over the pixel-to-pixel variations within any clean beam. No changes in spatial resolution results from this process.
\item Spectral index uncertainties have been thoroughly investigated. Apart from extreme values near the edges "edge-effect" %(which do not affect the weighted average $\alpha$ values) 
which should be ignored, there are effects of the primary beam corrections. Our tests on two pointing data show that results outside the half-power level of the primary beam model used in CASA need be treated with caution. 
The true error in $\alpha$ is typically underestimated by 20\% in the formal error maps. In regions of high signal-to-noise, they may even underestimate the error by a factor $\sim$5. The latter affects C-band more than L-band, since the galaxies do not extend beyond the half-power point of the primary beam for L-band. 

\item Galaxies with flatter $\alpha$ values at the centres often have indications of central activity. 
In some cases, such as NGC~2820, NGC~3556 and NGC~5775 and others, $\alpha$ is flatter at C-band than in L-band, suggesting a higher contribution from thermal emission at the higher frequency. 

%Two examples of large halo galaxies with unexpectedly flat spectral index are NGC~3556 and NGC~5775 at C-band, for which the data are well behaved, and no other instrumental or calibrational problems exist which could have affected the result. 

%\item We present weighted mean spectral indices for both bands and weightings (Table~\ref{tab:weightedalpha} as well as flux densitites with uncertainties (normally the calibration error of 2\%) (Table ~\ref{tab:sfr_etc}). 

\item Star formation rates were derived via WISE-WERGA images, and potential AGN contamination of the SFR results investigated. Many candidates may have starburst nuclei, with the exceptions of NGC~660, NGC~3735 and NGC~4388 who may indeed harbour an AGN. Lower limit SFR are calculated for these and used for the two latter galaxies.  
\item We scaled, rotated and ordered the L-band images of the CHANG-ES galaxies by SFR surface density in order to be able to compare the galaxies on equal footing and get a snap shot of potential correlations between SFR and halo size. We produced two maps, one scaled by distance (to 10 Mpc), the other by the 22\micron\ WISE-WERGA sizes, and the second one does reveal a trend of galaxies harbouring larger halos having higher SFR densities. 
\item The L-band images from the previous point were medianed to bring forth the median edge-on galaxy, which harbours a compelling halo. 
\end{enumerate}

%% Included in this acknowledgments section are examples of the
%% AASTeX hypertext markup commands. Use \url without the optional [HREF]
%% argument when you want to print the url directly in the text. Otherwise,
%% use either \url or \anchor, with the HREF as the first argument and the
%% text to be printed in the second.

\acknowledgments
The National Radio Astronomy Observatory is a facility of the National Science Foundation operated under cooperative agreement by Associated Universities inc.

T.W, J.I., A.M., P.S. and C.M. thank the VLA staff for excellent assistance for these observations and reductions during their respective visits to Soccorro. 

We are greatly indebted to T. Jarrett for running the WISE WERGA process on our sample and providing us with the resulting images.

The work at Ruhr-University Bochum has been supported by DFG through FOR1048. 

We acknowledge the support of the Computer Center of the Max-Planck Institute (RZG) in Garching, Germany for the use of archiving facilities.

We are grateful to A. Vladimirov, for graciously providing time at a Stanford computer cluster for some of our imaging. In extension, we acknowledge support via NASA grants NNX10AE78G and NNX13AC47G. 

This research has made use of the NASA/IPAC Extragalactic Database (NED) which is operated by the Jet Propulsion Laboratory, California Institute of Technology, under contract with the National Aeronautics and Space Administration.

SDSS and DSS2 images were used to create the optical images in panel d) of the appendix images. 
Funding for the Sloan Digital Sky Survey (SDSS) and SDSS-II has been provided by the Alfred P. Sloan Foundation, the Participating Institutions, the National Science Foundation, the U.S. Department of Energy, the National Aeronautics and Space Administration, the Japanese Monbukagakusho, and the Max Planck Society, and the Higher Education Funding Council for England. The SDSS Web site is http://www.sdss.org/.
The SDSS is managed by the Astrophysical Research Consortium (ARC) for the Participating Institutions. The Participating Institutions are the American Museum of Natural History, Astrophysical Institute Potsdam, University of Basel, University of Cambridge, Case Western Reserve University, The University of Chicago, Drexel University, Fermilab, the Institute for Advanced Study, the Japan Participation Group, The Johns Hopkins University, the Joint Institute for Nuclear Astrophysics, the Kavli Institute for Particle Astrophysics and Cosmology, the Korean Scientist Group, the Chinese Academy of Sciences (LAMOST), Los Alamos National Laboratory, the Max-Planck-Institute for Astronomy (MPIA), the Max-Planck-Institute for Astrophysics (MPA), New Mexico State University, Ohio State University, University of Pittsburgh, University of Portsmouth, Princeton University, the United States Naval Observatory, and the University of Washington.

The Digitized Sky Surveys were produced at the Space Telescope Science Institute under U.S. Government grant NAG W-2166. The images of these surveys are based on photographic data obtained using the Oschin Schmidt Telescope on Palomar Mountain and the UK Schmidt Telescope. The plates were processed into the present compressed digital form with the permission of these institutions.

This work is based [in part] on observations made with the Spitzer Space Telescope and has made use of the NASA/ IPAC Infrared Science Archive, which are operated by the Jet Propulsion Laboratory, California Institute of Technology under a contract with NASA.  This publication makes use of data products from the Wide-field Infrared Survey Explorer, which is a joint project of the University of California, Los Angeles, and the Jet Propulsion Laboratory/California Institute of Technology, funded by NASA.

This research made use of APLpy, an open-source plotting package for Python hosted at http://aplpy.github.com.


%% in parentheses, after the keyword, but they will not be verified.

{\it Facilities:} \facility{VLA}, \facility{Spitzer}.



\appendix

\section{A smorgasbord of results for each CHANG-ES galaxy}
The figures in this appendix display an assortment of the data products from the CHANG-ES project, described below. 
With the exception of the field panel (j) in the red frame, all panels are displayed in the exact same angular field of view, for better comparison. 
The 'apparent B-vectors' in panels (d), (e) and (f) below, are in reality E-vectors rotated by 90 degrees (see Sec.~\ref{sec:polarizationimaging}).

{\it First row}
\\
a) C-band Stokes I intensity map with a Briggs robust 0 weighting. The I contours are overlaid for clarity, at 3, 6, 12, 24, 48, 96, 192 times the rms noise of the map (rms values are listed Table~\ref{tab:imagingparametersC}).  
\\
b) C-band Stokes I intensity map with a uv-tapering applied onto the robust 0 weighting, yielding a beam approximately twice the size of that of the robust 0 weighting in C-band, overlaid with its contours at 3, 6, 12, 24, 48, 96, 192 times the rms noise. This map does not exist for galaxies NGC~4438 and NGC~4845. 
\\
c) L-band Stokes I intensity map, with contours at 3, 6, 12, 24, 48, 96, 192 times the L rms noise (see Table~\ref{tab:imagingparametersL} for rms values).

{\it Second row:}
\\
d) C-band Stokes I contours with uv-tapered weighting at 3 times the rms noise, with apparent B-vectors at the same weighting , overlaid on an optical image. The optical images were created using a combination of Sloan Digital Sky Survey (SDSS) g,r,i bands {\it or} Digitized Sky Survey 2 (DSS2) blue red and infrared bands, for the galaxies not available in SDSS. 
\\
e) C-band polarization map with apparent B-vectors and Stokes I contours at 3$\sigma$. 
\\
f) L-band polarization map with apparent B-vectors and Stokes I contours at 3$\sigma$. \\

{\it Third row:}
\\
g) Stokes I intensity map with 3 weightings merged (images a, b and c above). Image a, i.e. C-band with robust 0 weighting, is shown in dark blue, image (b), i.e. C-band with uv-tapering, is shown in green, and image c, L-band robust 0 weighting, is shown in white. Except for six galaxies for which no satisfactory uv-tapered L-band image has been obtained (see Table~\ref{tab:imagingparametersL} the background has been masked out using the L-band uv-tapered map (this map is not shown separately). 
\\
h) C-band spectral index map. Spectral indices have been cut off where formal errors are larger than 1.0, and then smoothed/averaged to the size of the beam (Section~\ref{sec:spectral_indices}). The colour bar of panel i) also applies to this panel. 
\\
i) L-band spectral index map. Spectral indices have been cut off where formal errors are larger than 1.0, and then smoothed/averaged to the size of the beam (Section~\ref{sec:spectral_indices}).  \\

{\it Fourth row:}
\\
j) The L-band robust 0 surrounding environment. The red frame indicates that this panel is displayed in a different field of view than the other panels. 
\\
k) Spectral index uncertainty map of (h) - formal error data. See Section~\ref{sec:alpha_errors} for more information on the uncertainties. Note that although the true error is higher than what is displayed in this map, the map gives an indication of which regions are to be treated with caution. The colour bar of panel (l) also applies to this panel.
\\
l) Spectral index uncertainty map of (h) - formal error data. See caption of panel (k) and Section~\ref{sec:alpha_errors} for more information on the uncertainties. 
%
%%%%%%%%%%%%%% IMAGES %%%%%%%%%%%%%%%
\newpage
\onecolumn
\begin{figure}[H]
\includegraphics[angle=0,width=18cm]{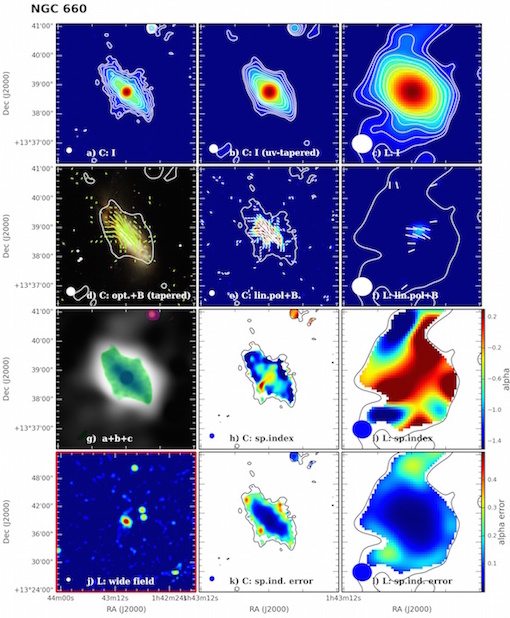}
\label{fig:n660}
\end{figure}
\newpage
\begin{figure}[H]
\includegraphics[angle=0,width=18cm]{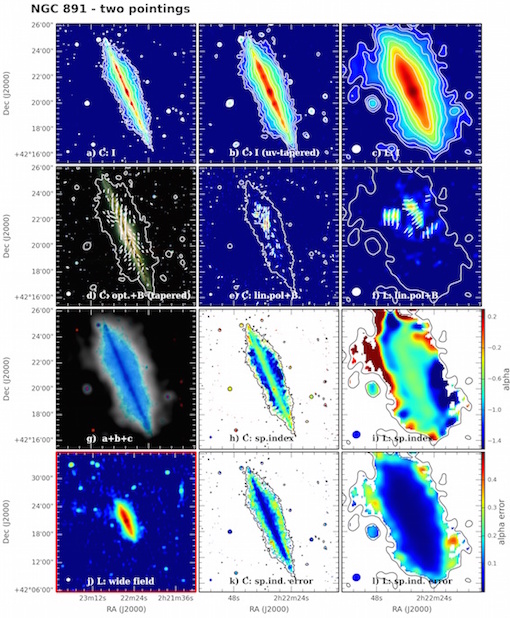}
\label{fig:n891}
\end{figure}
\newpage
\begin{figure}[H]
\includegraphics[angle=0,width=18cm]{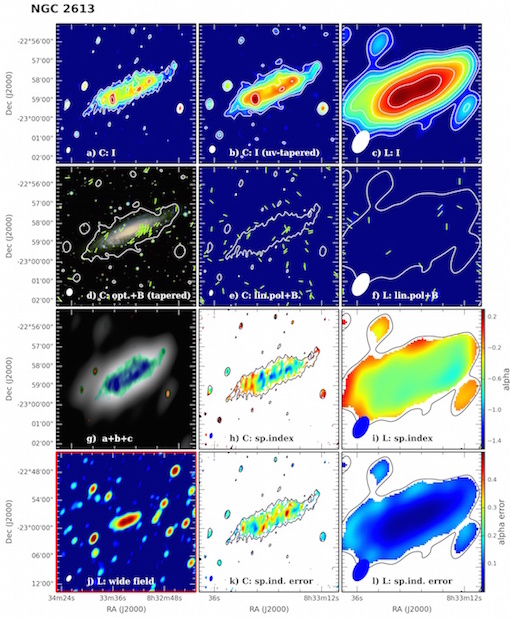}
\label{fig:n2613}
\end{figure}
\newpage
\begin{figure}[H]
\includegraphics[angle=0,width=18cm]{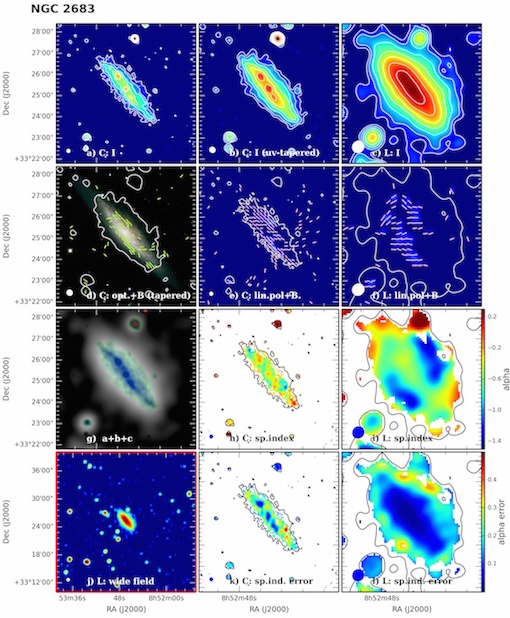}
\label{fig:n2683}
\end{figure}
\newpage
\begin{figure}[H]
\includegraphics[angle=0,width=18cm]{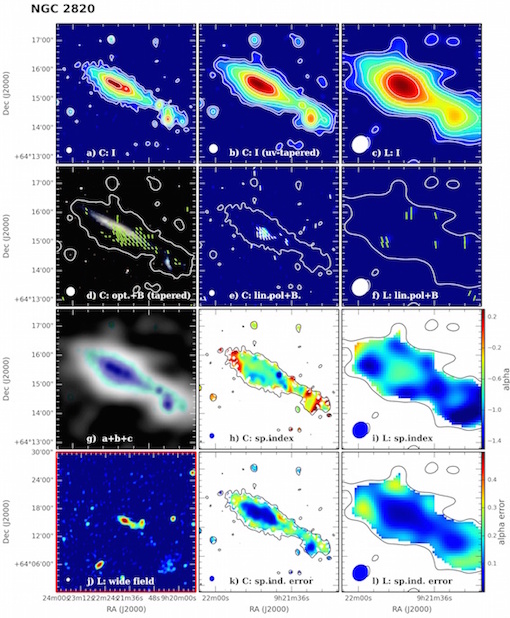}
\label{fig:n2820}
\end{figure}
\newpage
\begin{figure}[H]
\includegraphics[angle=0,width=18cm]{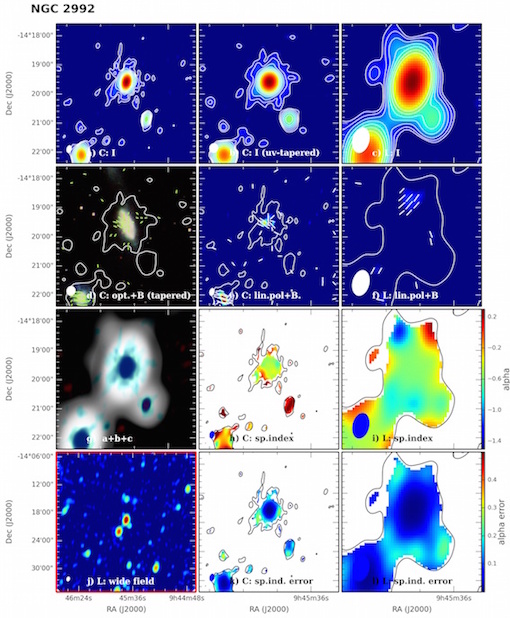}
\label{fig:n2992}
\end{figure}
\newpage
\begin{figure}[H]
\includegraphics[angle=0,width=18cm]{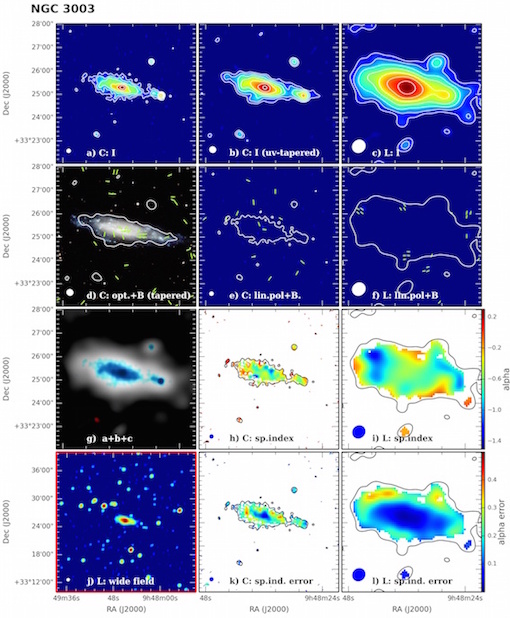}
\label{fig:n3003}
\end{figure}
\newpage
\begin{figure}[H]
\includegraphics[angle=0,width=18cm]{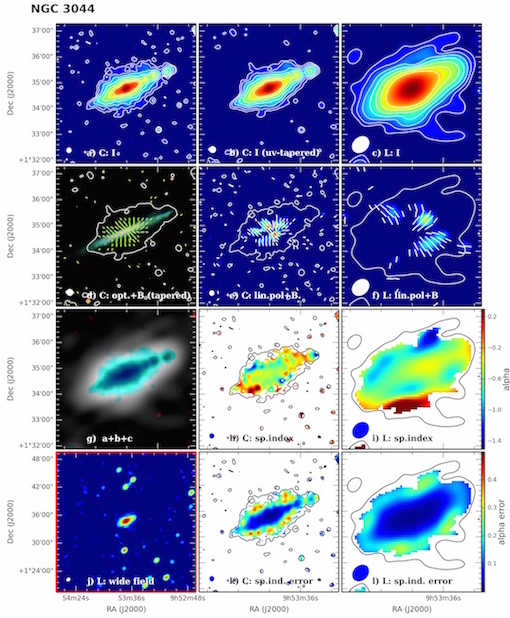}
\label{fig:n3044}
\end{figure}
\newpage
\begin{figure}[H]
\includegraphics[angle=0,width=18cm]{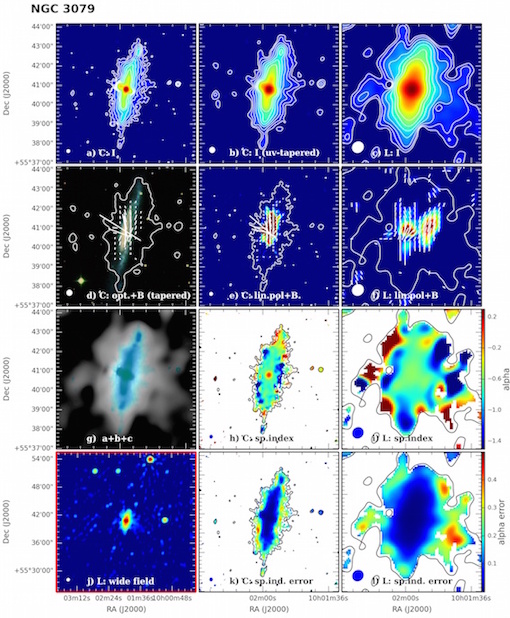}
\label{fig:n3079}
\end{figure}
\newpage
\begin{figure}[H]
\includegraphics[angle=0,width=18cm]{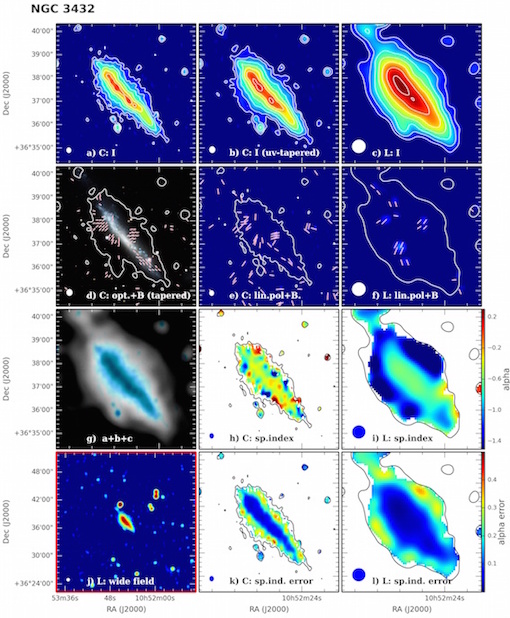}
\label{fig:n3432}
\end{figure}
\newpage
\begin{figure}[H]
\includegraphics[angle=0,width=18cm]{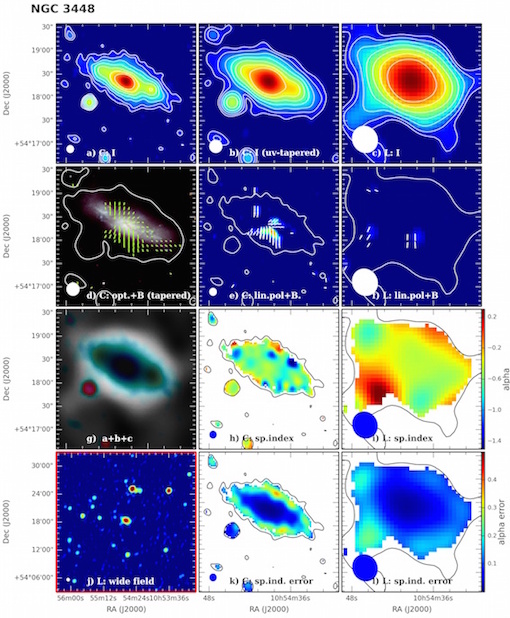}
\label{fig:n3448}
\end{figure}
\newpage
\begin{figure}[H]
\includegraphics[angle=0,width=18cm]{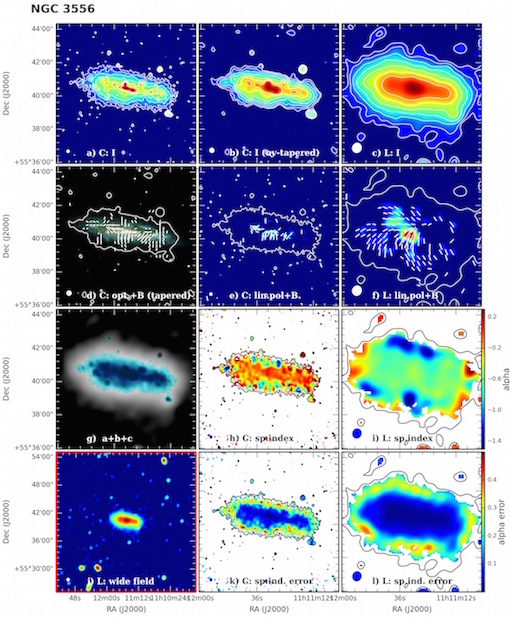}
\label{fig:n3556}
\end{figure}
\newpage
\begin{figure}[H]
\includegraphics[angle=0,width=18cm]{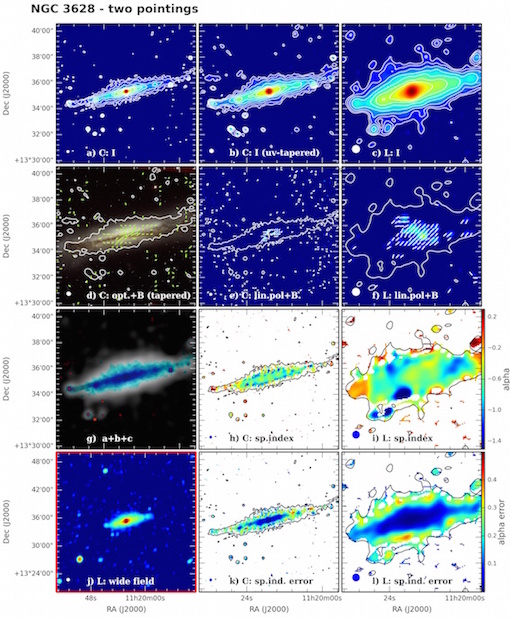}
\label{fig:n3628}
\end{figure}
\newpage
\begin{figure}[H]
\includegraphics[angle=0,width=18cm]{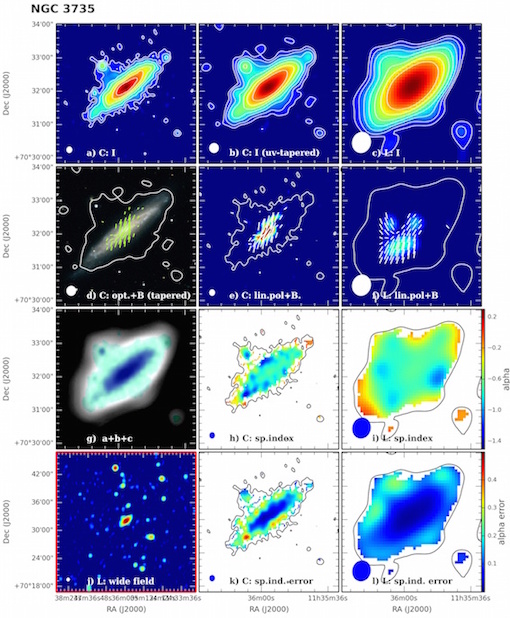}
\label{fig:n3735}
\end{figure}
\newpage
\begin{figure}[H]
\includegraphics[angle=0,width=18cm]{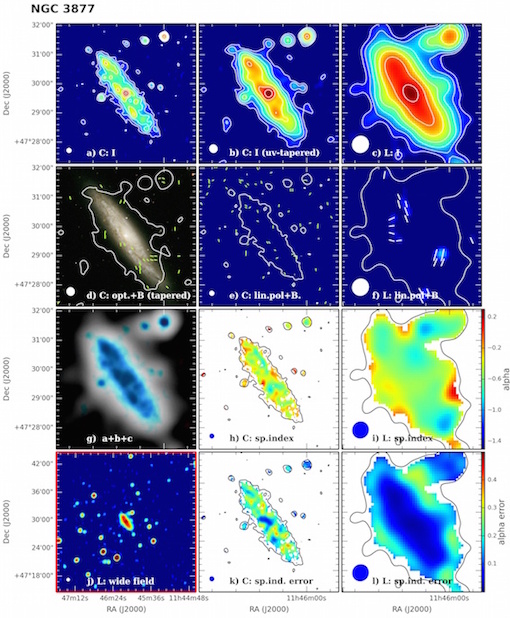}
\label{fig:n3877}
\end{figure}
\newpage
\begin{figure}[H]
\includegraphics[angle=0,width=18cm]{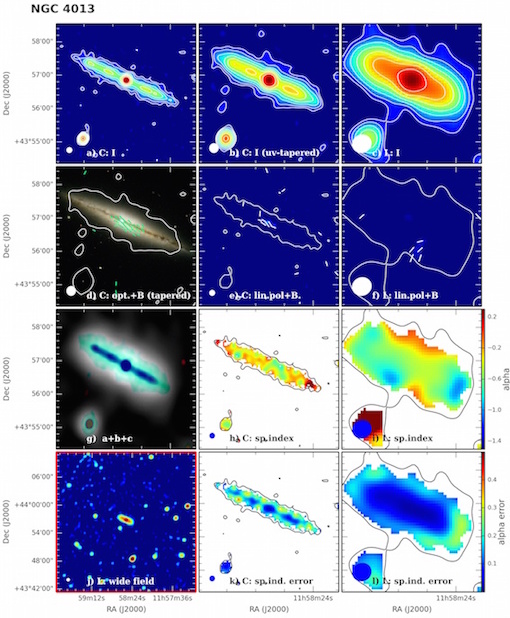}
\label{fig:n4013}
\end{figure}
\newpage
\begin{figure}[H]
\includegraphics[angle=0,width=18cm]{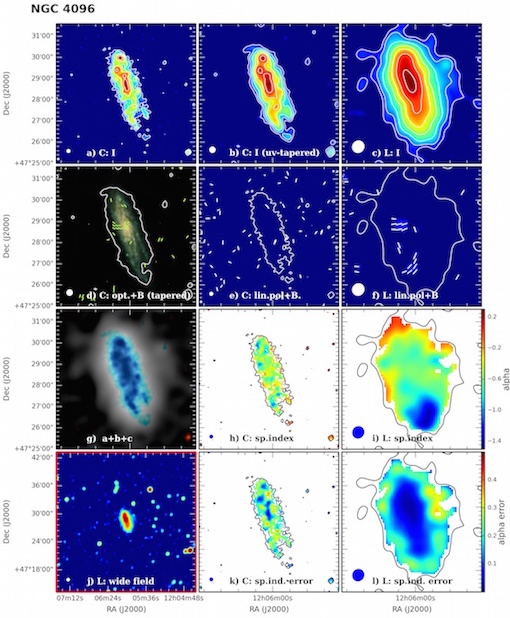}
\label{fig:n4096}
\end{figure}
\newpage
\begin{figure}[H]
\includegraphics[angle=0,width=18cm]{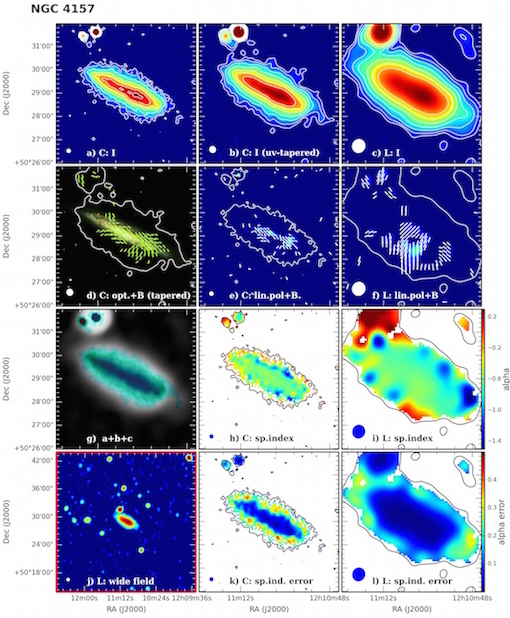}
\label{fig:n4157}
\end{figure}
\newpage
\begin{figure}[H]
\includegraphics[angle=0,width=18cm]{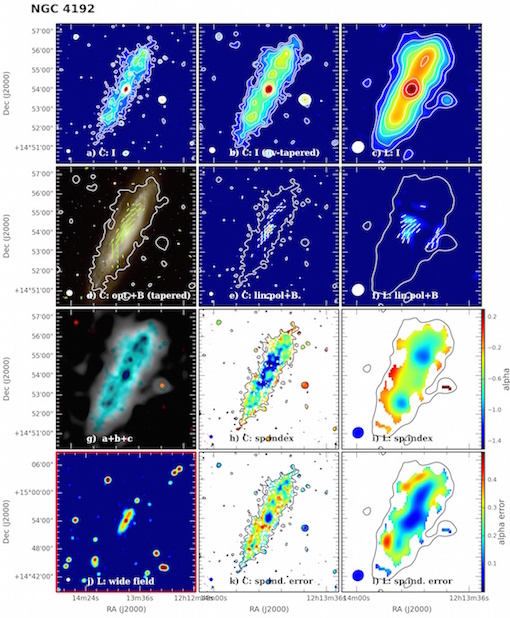}
\label{fig:n4192}
\end{figure}
\newpage
\begin{figure}[H]
\includegraphics[angle=0,width=18cm]{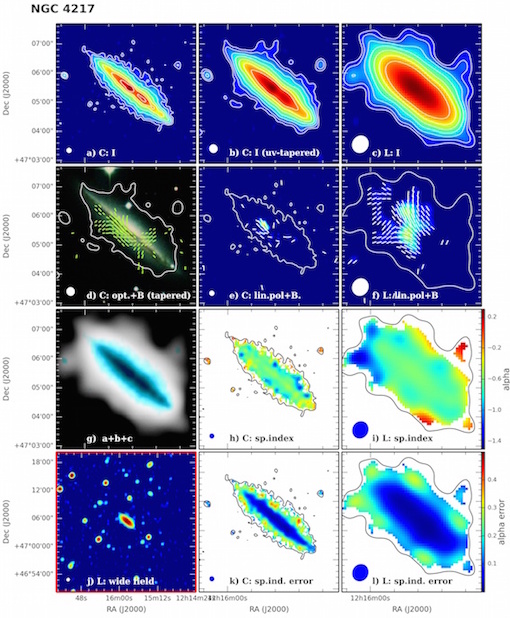}
\label{fig:n4217}
\end{figure}
\newpage
\begin{figure}[H]
\includegraphics[angle=0,width=18cm]{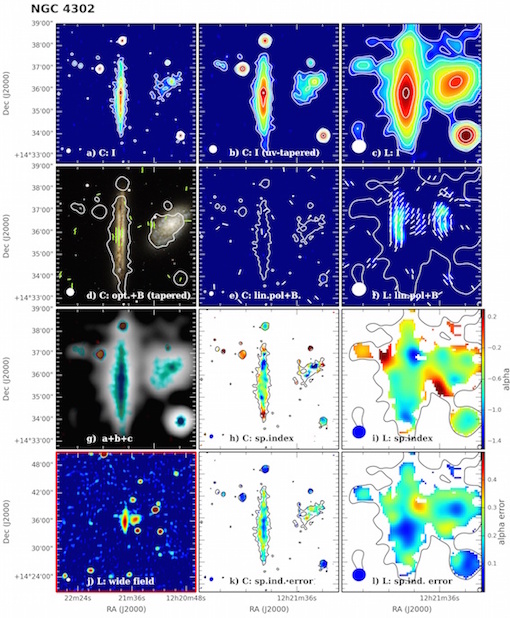}
\label{fig:n4302}
\end{figure}
\newpage
\begin{figure}[H]
\includegraphics[angle=0,width=18cm]{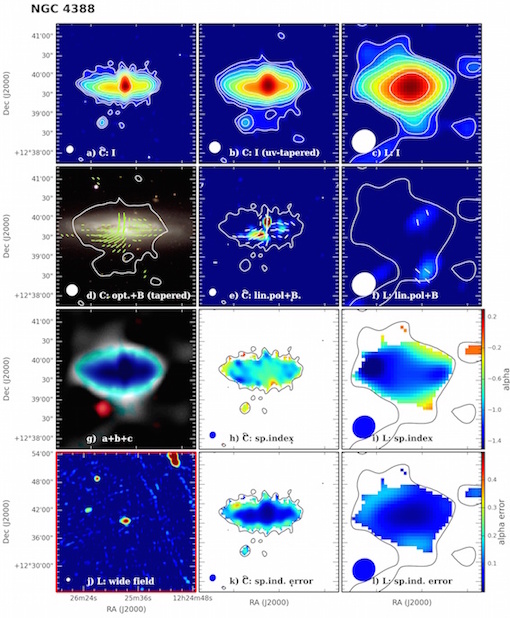}
\label{fig:n4388}
\end{figure}
\newpage
\begin{figure}[H]
\includegraphics[angle=0,width=18cm]{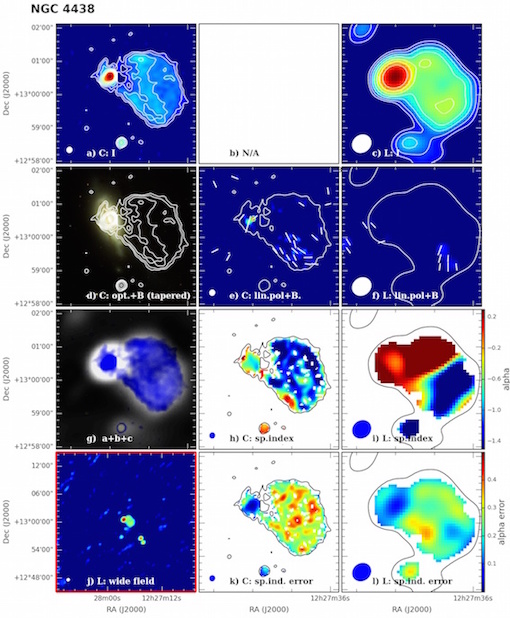}
\label{fig:n4438}
\end{figure}
\newpage
\begin{figure}[H]
\includegraphics[angle=0,width=18cm]{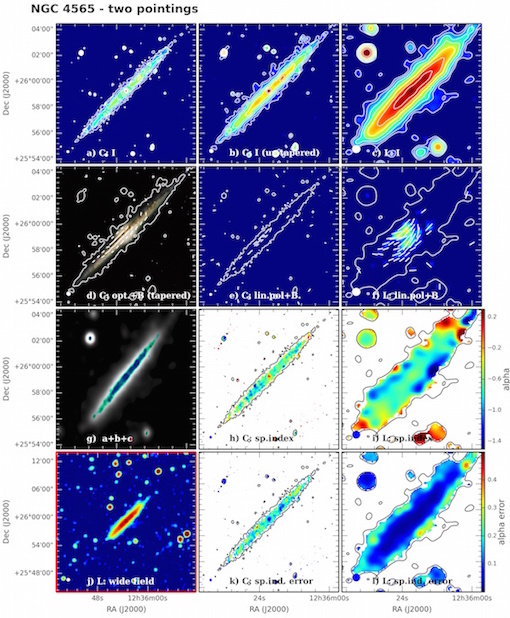}
\label{fig:n4565}
\end{figure}
\newpage
\begin{figure}[H]
\includegraphics[angle=0,width=18cm]{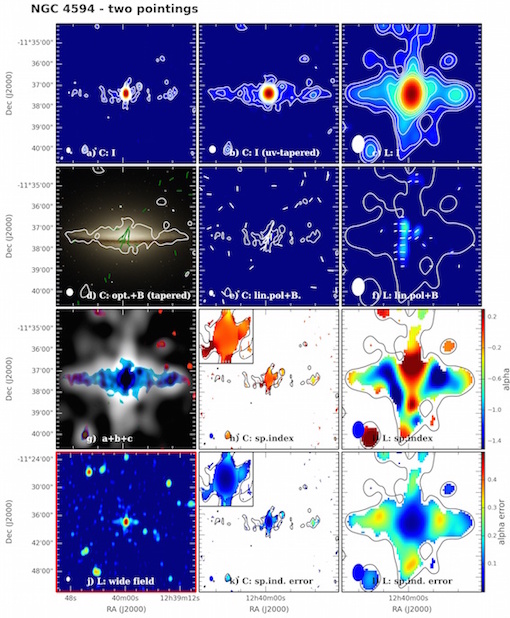}
\label{fig:n4594}
\end{figure}
\newpage
\begin{figure}[H]
\includegraphics[angle=0,width=18cm]{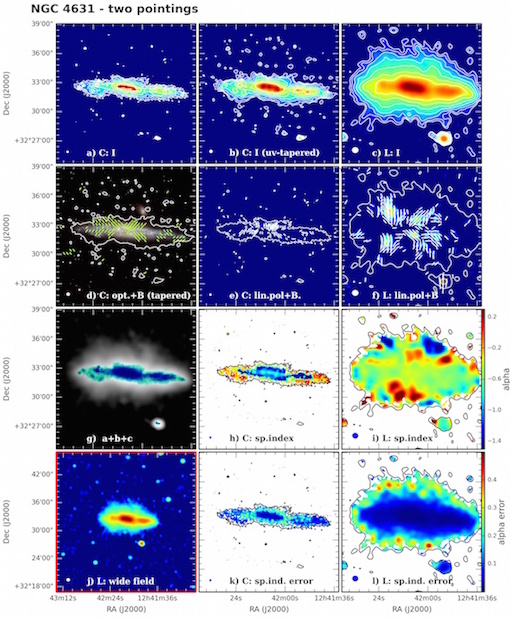}
\label{fig:n4631}
\end{figure}
\newpage
\begin{figure}[H]
\includegraphics[angle=0,width=18cm]{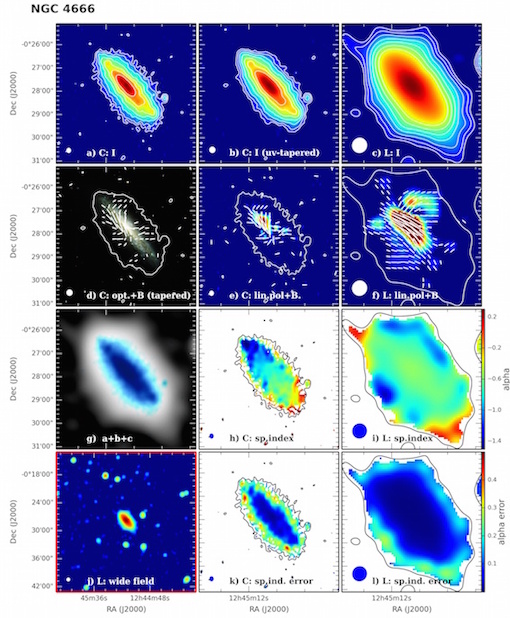}
\label{fig:n4666}
\end{figure}
\newpage
\begin{figure}[H]
\includegraphics[angle=0,width=18cm]{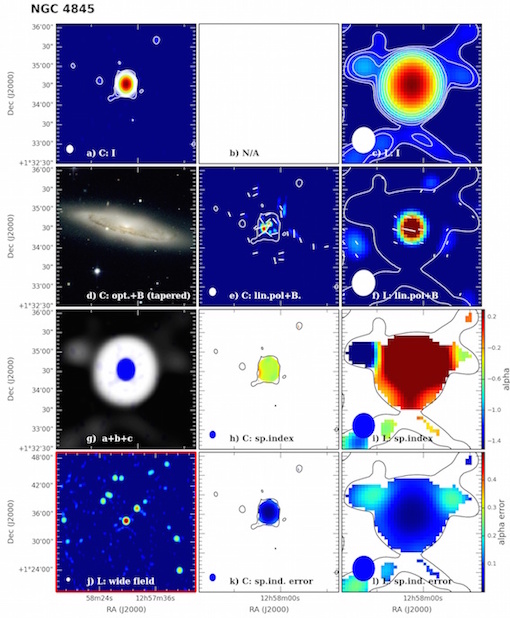}
\label{fig:n4845}
\end{figure}
\newpage
\begin{figure}[H]
\includegraphics[angle=0,width=18cm]{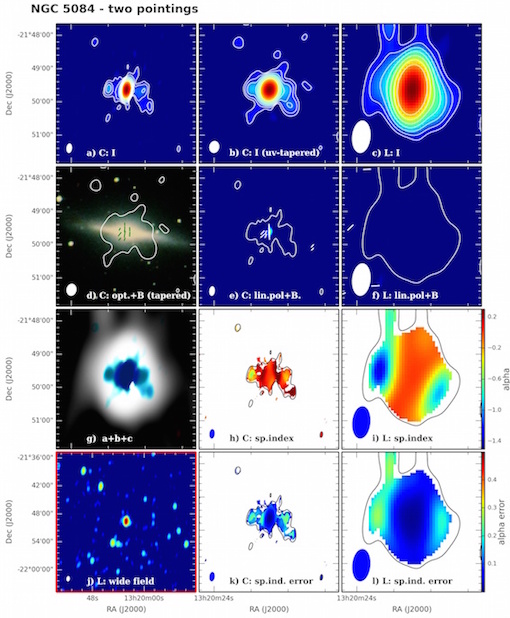}
\label{fig:n5084}
\end{figure}
\newpage
\begin{figure}[H]
\includegraphics[angle=0,width=18cm]{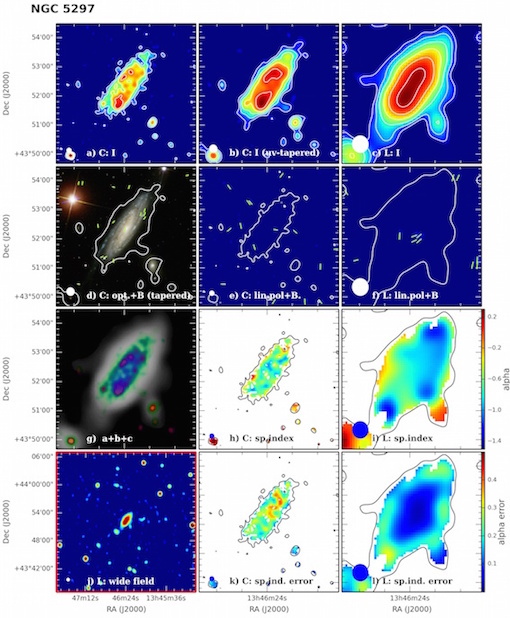}
\label{fig:n5297}
\end{figure}
\newpage
\begin{figure}[H]
\includegraphics[angle=0,width=18cm]{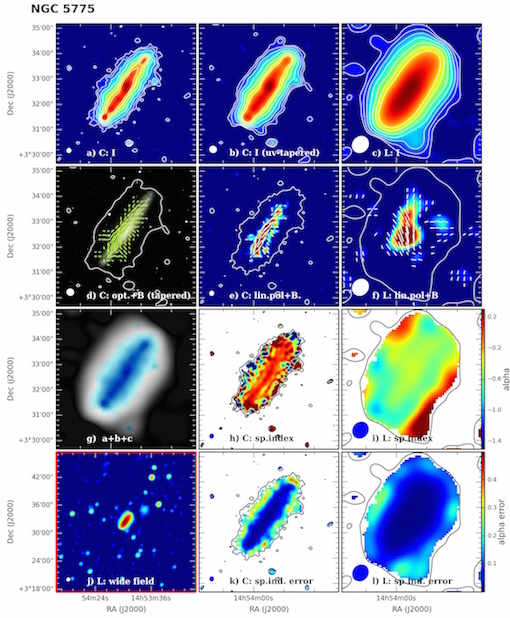}
\label{fig:n5775}
\end{figure}
\newpage
\begin{figure}[H]
\includegraphics[angle=0,width=18cm]{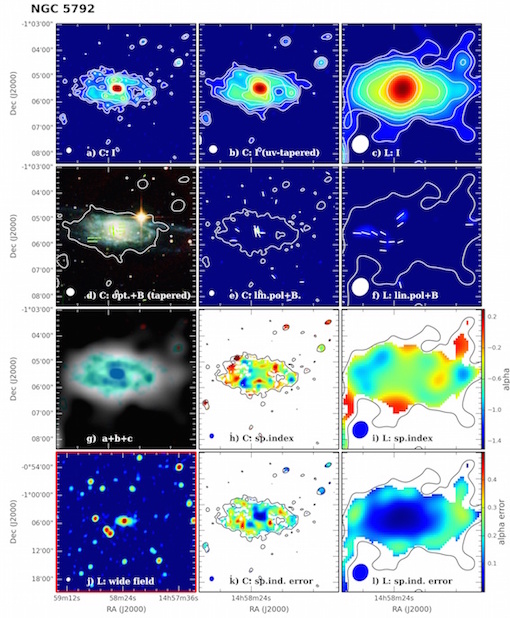}
\label{fig:n5792}
\end{figure}
\newpage
\begin{figure}[H]
\includegraphics[angle=0,width=18cm]{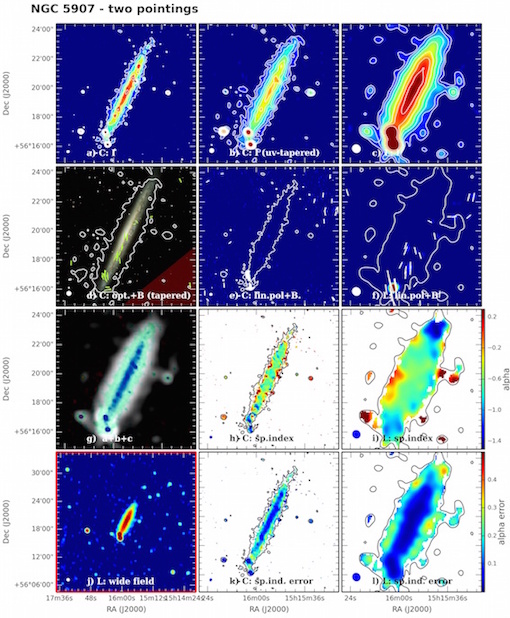}
\label{fig:n5907}
\end{figure}
\newpage
\begin{figure}[H]
\includegraphics[angle=0,width=18cm]{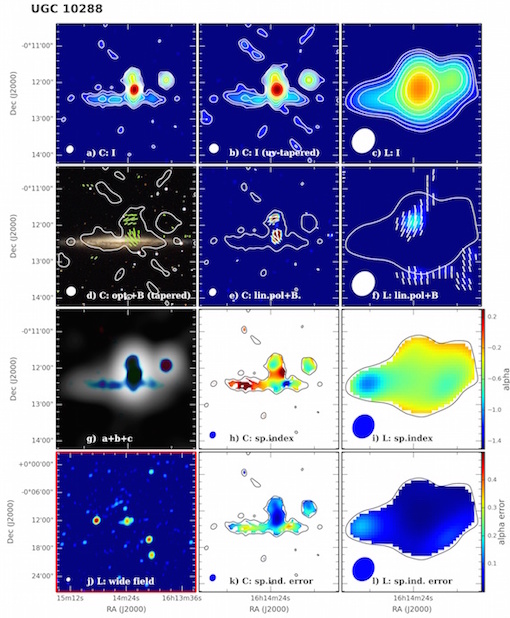}
\label{fig:u10288}
\end{figure}

%plus the rest!!




\clearpage
%% the file, uncomment the \plotone, \plottwo, and
%% If you need a layout that cannot be achieved with \plotone or
%% \plottwo, you can invoke the graphicx package directly with the
%% \includegraphics command or use \plotfiddle. For more information,
%% please see the tutorial on "Using Electronic Art with AASTeX" in the
%% documentation section at the AASTeX Web site, http://aastex.aas.org/
%%
%% The examples below also include sample markup for submission of
%% supplemental electronic materials. As always, be sure to check
%% the instructions to authors for the journal you are submitting to
%% for specific submissions guidelines as they vary from

%% This example uses \plotone to include an EPS file scaled to
%% 80% of its natural size with \epsscale. Its caption
%% has been written to indicate that additional figure parts will be
%% available in the electronic journal.

%% Here we use \plottwo to present two versions of the same figure,
%% one in black and white for print the other in RGB color
%% for online presentation. Note that the caption indicates
%% that a color version of the figure will be available online.
%%
%\begin{figure}
%\plottwo{f2.eps}{f2_color.eps}
%\caption{A panel taken from Figure 2 of \citet{rudnick03}. 
%See the electronic edition of the Journal for a color version of this figure.\label{fig2}}
%\end{figure}

%%%%%%%%%%%%%% FIGURES %%%%%%%%%%%%%%%%
\newpage
\onecolumn
\begin{figure}[!tc]
\centering
\begin{tabular}{@{}cc@{}}
	\includegraphics[width=.46\textwidth]{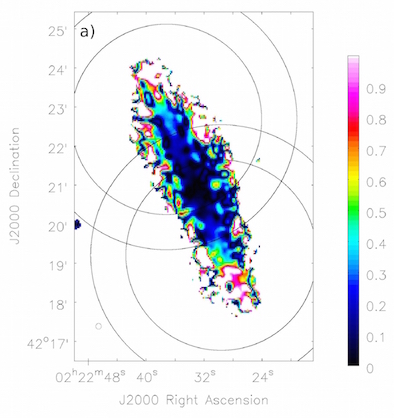} &
	\includegraphics[width=.46\textwidth]{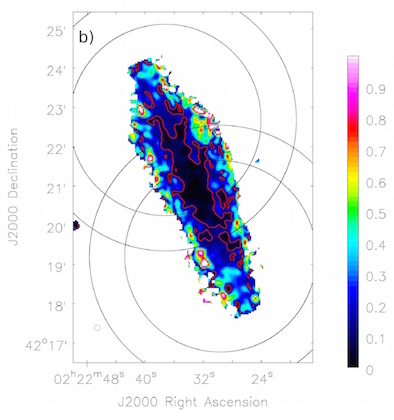} \\
	\includegraphics[width=.46\textwidth]{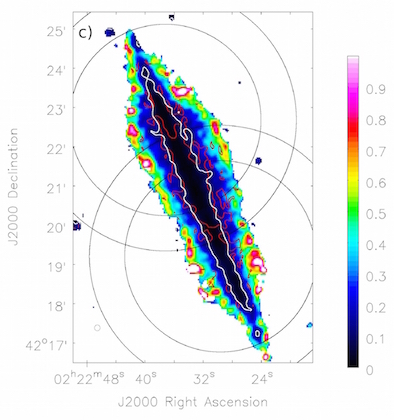} &
	\includegraphics[width=.46\textwidth]{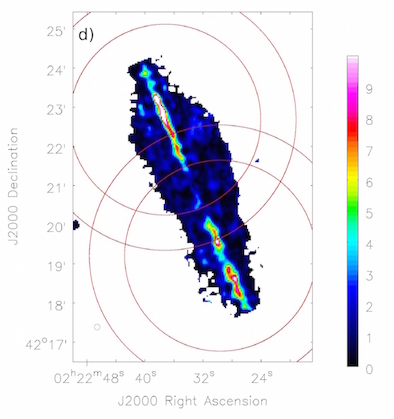} \\
\end{tabular}
\caption{a) Absolute difference between the PB-corrected spectral index measurements in the two C-band pointings of NGC~891. There is an increase in difference with distance from the midpoint between the pointing centres, even in regions with high S/N. b) Error of the PB-weighted average of the PB-corrected spectral index measurements in the two pointings. The red contour is placed at an error of 0.1. c) Spectral index error map determined by the ms-mfs clean algorithm (same as panel (k) of the appendix figure of NGC~891, but without the cutoff at $\Delta\,\alpha\,>\,1.0$). Contours represent an error of 0.1 in this map (white) and in the map shown in panel b) (red). d) Ratio of the two different error maps (panel b) divided by panel c)). The ratio is mostly close to unity, except for those parts of the disk lying outside the 0.7 PB level of either pointing.
In each panel, the two PB circular contours for each pointing are placed at the 0.5 and 0.7 level.
}
\label{fig:Philipfig}
\end{figure}

%\newpage
%\onecolumn
%\begin{figure}[!tc]
%\includegraphics[angle=0,width=12cm]{f37.eps}
%\caption{Absolute difference between the PB-corrected spectral index measurements in the two C-band pointings of NGC~891. Here the reference frequency has been shifted from 6.0 GHz to 6.6GHz, thereby artificially making the PB pattern narrower, and there is a better agreement between pointings than in the previous figure.
%}
%\label{fig:philip_spixN891_freqshift}
%\end{figure}

\newpage
\onecolumn
\begin{figure}[!t]
\includegraphics[angle=0, width=16cm]{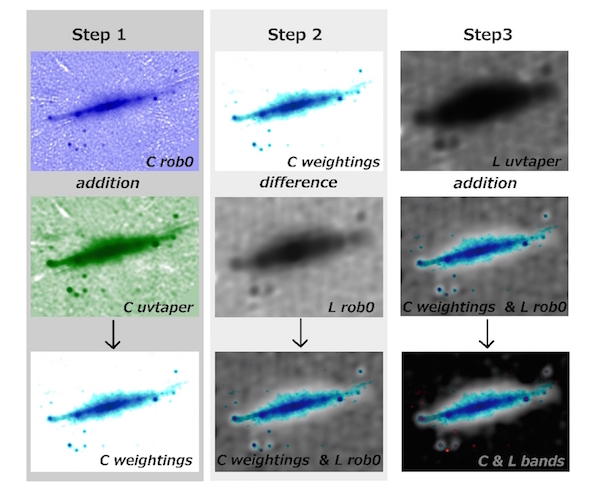}
\caption{Construction of combined weightings and bands in panel (g) of the figures in Appendix.   
The layering procedure in the Gnu Image Manipulation Program (GIMP) is used to combine the available weightings of the C and L band data. ("rob0" stands for the robust 0 weighting and "uvtap" is the uv-tapered weighting.) The algorithmic mode applied to the top layer is listed under the upper image in each column.  The resulting combination is displayed in the bottom row.  The left hand column represents the first 2 layers that are combined, i.e. inverted-intensity and colourized versions of the two weightings of the C band data.  The second step, represented in the middle column, combines the inverted-intensity higher resolution L band data with the result of Step 1. The right hand column shows how the inverted-intensity, lower resolution L band data are applied to  the result of Step 2 in order to mask out confusing structures in the off-target "background".  The bottom image in the right hand column is presented as panel (g).
}
\label{fig:jayanneflowchart}
\end{figure}

\newpage
\onecolumn
\begin{figure}[!t]
\includegraphics[angle=0,width=12cm]{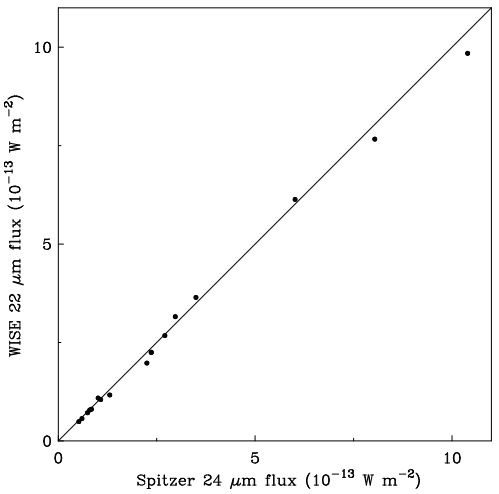}
\caption{WISE 22 $\mu$m vs. Spitzer 24\micron\ flux ($\nu F_\nu$) for the 16 CHANG-ES galaxies with archived Spitzer 24 $\mu$m images.  The line is not a fit to the data but shows equality of the two fluxes.
}
\label{fig:spitzerwise}
\end{figure}

\newpage
\onecolumn
\begin{figure}[!t]
\includegraphics[angle=0,height=18cm]{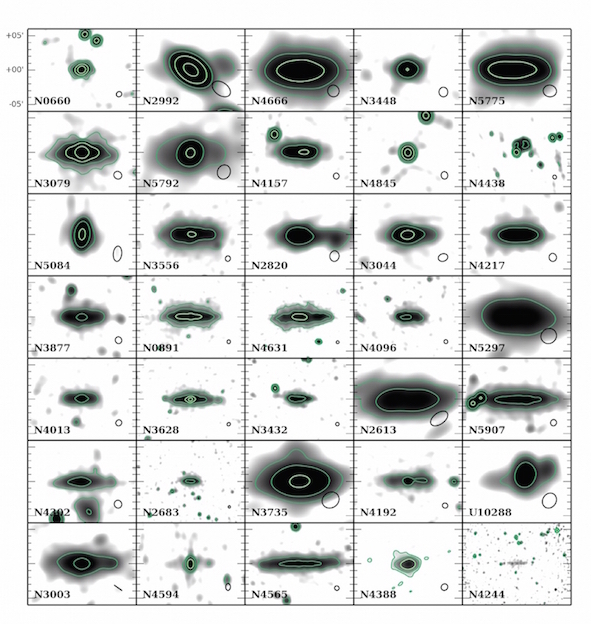}
\caption{The CHANG-ES galaxies L-band data, rotated and scaled by size to a distance of 10 MPc, ordered by SFR density. Highest SFR density at top left, decreasing by row, left to right. Contours shown are at levels of 1, 5, 25 and 125 mJy/beam. 5\arcmin\ corresponds to 14.5 kpc at this scale.} 
\label{fig:megamap1b}
\end{figure}

\newpage
\onecolumn
\begin{figure}[!t]
\includegraphics[angle=0,height=18cm]{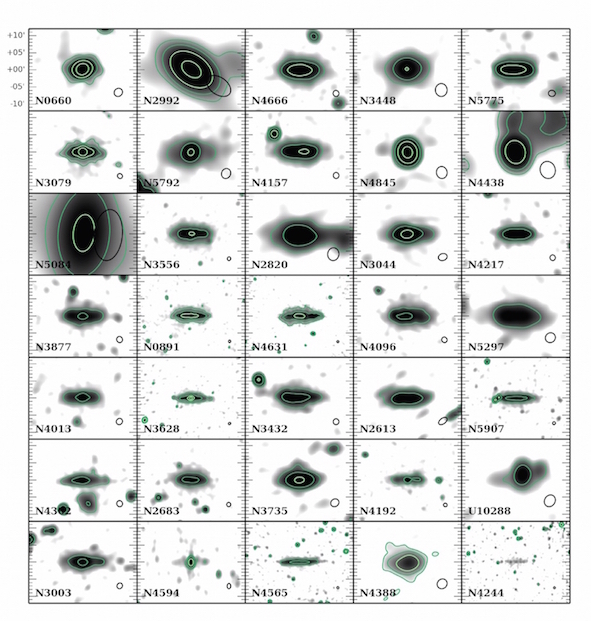}
\caption{The CHANG-ES galaxies L-band data, rotated, and scaled to the 22 micron WISE-WERGA size of NGC~4244, ordered by SFR surface density. Highest SFR surface density at top left, decreasing by row, left to right. Contours shown are at levels of 1, 5, 25 and 125 mJy/beam. }
\label{fig:megamap3}
\end{figure}

\newpage
\onecolumn
\begin{figure}[!t]
\includegraphics[angle=0,width=17cm]{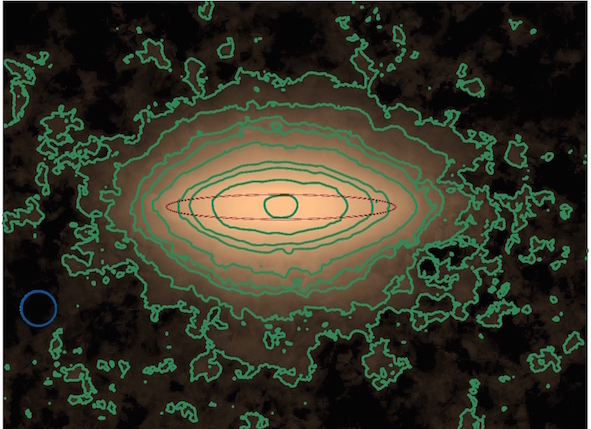}
\caption{The median edge-on spiral galaxy in L-band, made from stacking 30 of the galaxies in Fig.~\ref{fig:megamap3}. The red ellipse is a sample 22\micron\ contour that corresponds to the scaling of the radio data and thus represents the disk radial extent. The beam shown is the {\it average} beam of the 30 galaxies.}
\label{fig:mediangalaxy}
\end{figure}

%% Tables should be submitted one per page, so put a \clearpage before

%% Two options are available to the author for producing tables:  the
%% deluxetable environment provided by the AASTeX package or the LaTeX
%% table environment.  Use of deluxetable is preferred.
%%

%% Three table samples follow, two marked up in the deluxetable environment,
%% one marked up as a LaTeX table.

%% In this first example, note that the \tabletypesize{}
%% command has been used to reduce the font size of the table.
%% We also use the \rotate command to rotate the table to
%% landscape orientation since it is very wide even at the
%% Note also that the \label command needs to be placed

%% This table also includes a table comment indicating that the full
%% version will be available in machine-readable format in the electronic
\clearpage

%%%%%%%%%%%%% TABLE 1 - THE OBSERVATIONS %%%%%%%%%%%%%%%
\begin{deluxetable}{cccccccccc}
\tabletypesize{\scriptsize}
\rotate
\tablecaption{Observations \label{tab:observations}}
\tablewidth{0pt}
\tablehead{
\colhead{Galaxy\tablenotemark{a}} & \colhead{RA} & \colhead{dec} & Distance\tablenotemark{b} (Mpc) & Band & \colhead{Date \it{(yymmdd)}} & \colhead{SB ID} & \colhead{prim. cal.}\tablenotemark{c} & \colhead{zero. pol. cal.\tablenotemark{c}} & \colhead{sec. cal.}
}
\startdata
N 660	&  01h43m02.s40  & +13d38m42.s2  &12.3 & C  & 111209  & 5804098  & 3C48  & 3C84  & J0204+1514   \\  
	&		 &		 &      & C  & 111218  & 5805266  & 3C48  & 3C84  & J0204+1514   \\
        &		 &		 &      & L  & 111219  & 6621021  & 3C48  & 3C84  & J0204+1514   \\
	& 		 &		 &      & L  & 130317  & 6619959  & 3C48  & 3C84  & J0204+1514   \\
N 891*	&  02h22m33.s41  & +42d20m56.s9  & 9.1* & C  & 111209  & 5804098  & 3C48  & 3C84  & J0230+4032   \\
	&		 &		 &      & C  & 111218  & 5805266  & 3C48  & 3C84  & J0230+4032   \\
	& 		 &		 &      & L  & 111219  & 6621021  & 3C48  & 3C84  & J0314+4314   \\
	&		 &		 &      & L  & 130317  & 6619959  & 3C48  & 3C84  & J0314+4314   \\
N 2613	&  08h33m22.s84  & -22d58m25.s2  & 23.4 & C  & 111213  & 4806011  & 3C286 & OQ208 & J0837-1951   \\
	& 		 &		 &      & L  & 130317  & 6619959  & 3C48  & 3C84  & J0853-2047   \\
	&		 &		 &      & L  & 111221  & 4807896  & 3C286 & OQ208 & J0853-2047   \\
N 2683	&  08h52m41.s35  & +33d25m18.s5  & 6.27 & C  & 111213  & 4806011  & 3C286 & OQ208 & J0837+2454   \\
	& 		 &		 &      & L  & 111221  & 4807896  & 3C286 & OQ208 & J0909+4253   \\
N 2820	&  09h21m45.s58  & +64d15m28.s6  & 26.5 & C  & 111217  & 4809751  & 3C286 & OQ208 & J0921+6215   \\
	& 		 &		 &      & L  & 111218  & 4812474  & 3C286 & OQ208 & J0949+6614   \\
N 2992	&  09h45m42.s00  & -14d19m35.s0  & 34   & C  & 111213  & 4806011  & 3C286 & OQ208 & J0943-0819   \\
	& 		 &		 &      & L  & 111221  & 4807896  & 3C286 & OQ208 & J0943-0819   \\
N 3003	&  09h48m36.s05  & +33d25m17.s4  & 25.4 & C  & 111213  & 4806011  & 3C286 & OQ208 & J0958+3224   \\
	& 		 &		 &      & L  & 111221  & 4807896  & 3C286 & OQ208 & J0958+3224   \\
N 3044	&  09h53m40.s88  & +01d34m46.s7  & 20.3 & C  & 111213  & 4806011  & 3C286 & OQ208 & J0925+0019   \\
	&		 &		 &      & L  & 111221  & 4807896  & 3C286 & OQ208 & J1007-0207   \\
N 3079	&  10h01m57.s80  & +55d40m47.s3  & 20.6 & C  & 111217  & 4809751  & 3C286 & OQ208 & J1035+5628   \\
	&		 &		 &      & L  & 111218  & 4812474  & 3C286 & OQ208 & J1035+5628   \\
N 3432	&  10h52m31.s13  & +36d37m07.s6  & 9.42 & C  & 111217  & 4809751  & 3C286 & OQ208 & J1104+3812   \\
	&		 &		 &      & L  & 111218  & 4812474  & 3C286 & OQ208 & J1006+3454   \\
N 3448	&  10h54m39.s24  & +54d18m18.s8  & 24.5 & C  & 111217  & 4809751  & 3C286 & OQ208 & J1035+5628   \\
	&		 &		 &      & L  & 111218  & 4812474  & 3C286 & OQ208 & J1035+5628   \\
N 3556	&  11h11m30.s97  & +55d40m26.s8  & 14.09& C  & 111217  & 4809751  & 3C286 & OQ208 & J1035+5628   \\
	&		 &		 &      & L  & 111218  & 4812474  & 3C286 & OQ208 & J1035+5628   \\
N 3628*	&  11h20m17.s01  & +13d35m22.s9  & 8.5  & C  & 111213  & 4806011  & 3C286 & OQ208 & J1120+1420   \\
	&		 &		 &      & L  & 111221  & 4807896  & 3C286 & OQ208 & J1120+1420   \\
N 3735	&  11h35m57.s30  & +70d32m08.s1  & 42   & C  & 111217  & 4809751  & 3C286 & OQ208 & J1056+7011   \\
	&		 &		 &      & L  & 111218  & 4812474  & 3C286 & OQ208 & J1206+6413   \\
N 3877	&  11h46m07.s80  & +47d29m41.s2  & 17.7 & C  & 111227  & 5062561  & 3C286 & OQ208 & J1219+4829   \\
	&		 &		 &      & L  & 111218  & 4812474  & 3C286 & OQ208 & J1219+4829   \\
N 4013	&  11h58m31.s38  & +43d56m47.s7  & 16   & C  & 111227  & 5062561  & 3C286 & OQ208 & J1146+3958   \\
	&		 &		 &      & L  & 111218  & 4812474  & 3C286 & OQ208 & J1146+3958   \\
N 4096	&  12h06m01.s13  & +47d28m42.s4  & 10.32& C  & 111227  & 5062561  & 3C286 & OQ208 & J1146+3958   \\
	&		 &		 &      & L  & 111218  & 4812474  & 3C286 & OQ208 & J1146+3958   \\
N 4157	&  12h11m04.s37  & +50d29m04.s8  & 15.6 & C  & 111227  & 5062561  & 3C286 & OQ208 & J1219+4829   \\
	&		 &		 &      & L  & 111218  & 4812474  & 3C286 & OQ208 & J1219+4829   \\
N 4192	&  12h13m48.s29  & +14d54m01.s2  & 13.55V& C  & 111229  & 4809749  & 3C286 & OQ208 & J1239+0730   \\
	&		 &		 &      & L  & 111221  & 4807896  & 3C286 & OQ208 & J1254+1141   \\
	&		 &		 &      & L  & 111230  & 4812476  & 3C286 & OQ208 & J1254+1141   \\
N 4217	&  12h15m50.s90  & +47d05m30.s4  & 20.6 & C  & 111227  & 5062561  & 3C286 & OQ208 & J1219+4829   \\
	&		 &		 &      & L  & 111218  & 4812474  & 3C286 & OQ208 & J1219+4829   \\
N 4244*	&  12h17m29.s66  & +37d48m25.s6  & 4.4* & C  & 111217  & 4809751  & 3C286 & OQ208 & J1227+3635   \\
	&		 &		 &      & L  & 111218  & 4812474  & 3C286 & OQ208 & J1227+3635   \\
N 4302	&  12h21m42.s48  & +14d35m53.s9  & 19.41V& C  & 111219  & 5062559  & 3C286 & OQ208 & J1254+1141   \\
	&		 &		 &      & L  & 111221  & 4807896  & 3C286 & OQ208 & J1254+1141   \\
	&		 &		 &      & L  & 111230  & 4812476  & 3C286 & OQ208 & J1254+1141   \\
N 4388	&  12h25m46.s75  & +12d39m43.s5  & 16.6V & C  & 111219  & 5062559  & 3C286 & OQ208 & J1254+1141   \\
	&		 &		 &      & L  & 111221  & 4807896  & 3C286 & OQ208 & J1254+1141   \\
	&		 &		 &      & L  & 111230  & 4812476  & 3C286 & OQ208 & J1254+1141   \\
N 4438	&  12h27m45.s59  & +13d00m31.s8  & 10.39V& C  & 111219  & 5062559  & 3C286 & OQ208 & J1254+1141   \\
	&		 &		 &      & L  & 111221  & 4807896  & 3C286 & OQ208 & J1254+1141   \\
	&		 &		 &      & L  & 111230  & 4812476  & 3C286 & OQ208 & J1254+1141   \\
N 4565*	&  12h36m20.s78  & +25d59m15.s6  & 11.9*& C  & 111229  & 4809749  & 3C286 & OQ208 & J1221+2813   \\
	&		 &		 &      & L  & 111230  & 4812476  & 3C286 & OQ208 & J1221+2813   \\
N 4594*	&  12h39m59.s43  & -11d37m23.s0  & 12.7 & C  & 111219  & 5062559  & 3C286 & OQ208 & J1246-0730   \\
	&		 &		 &      & L  & 111230  & 4812476  & 3C286 & OQ208 & J1246-0730   \\
N 4631*	&  12h42m08.s01  & +32d32m29.s4  & 7.4* & C  & 111229  & 4809749  & 3C286 & OQ208 & J1310+3220   \\
	&		 &		 &      & L  & 111230  & 4812476  & 3C286 & OQ208 & J1221+2813   \\
N 4666	&  12h45m08.s59  & -00d27m42.s8  & 27.5 & C  & 111219  & 5062559  & 3C286 & OQ208 & J1246-0730   \\
	&		 &		 &      & L  & 111230  & 4812476  & 3C286 & OQ208 & J1246-0730   \\
N 4845	&  12h58m01.s19  & +01d34m33.s0  & 16.98V& C  & 111219  & 5062559  & 3C286 & OQ208 & J1246-0730   \\
	&		 &		 &      & L  & 111230  & 4812476  & 3C286 & OQ208 & J1246-0730   \\
N 5084*	&  13h20m16.s92  & -21d49m39.s3  & 23.4 & C  & 111229  & 4809749  & 3C286 & OQ208 & J1248-1959   \\
	&		 &		 &      & C  & 111213  & 4806011  & 3C286 & OQ208 & J1248-1959   \\
	&		 &		 &      & C  & 111210  & 5062309  & 3C286 & OQ208 & J1248-1959   \\
	&		 &		 &      & L  & 111230  & 4812476  & 3C286 & OQ208 & J1248-1959   \\
N 5297	&  13h46m23.s68  & +43d52m20.s5  & 40.4 & C  & 111227  & 5062561  & 3C286 & OQ208 & J1327+4326   \\
	&		 &		 &      & L  & 111218  & 4812474  & 3C286 & OQ208 & J1357+4353   \\
N 5775	&  14h53m58.s00  & +03d32m40.s1  & 28.9 & C  & 111210  & 5062309  & 3C286 & OQ208 & J1445+0958   \\
	&		 &		 &      & L  & 111230  & 4812476  & 3C286 & OQ208 & J1445+0958   \\
N 5792	&  14h58m22.s71  & -01d05m27.s9  & 31.7 & C  & 111210  & 5062309  & 3C286 & OQ208 & J1505+0326   \\
	&		 &		 &      & L  & 111230  & 4812476  & 3C286 & OQ208 & J1510-0543   \\
N 5907*	&  15h15m53.s77  & +56d19m43.s6  & 16.8*& C  & 111227  & 5062561  & 3C286 & OQ208 & J1438+6211   \\
	&		 &		 &      & L  & 111230  & 4812476  & 3C286 & OQ208 & J1438+6211   \\
U 10288	&  16h14m24.s80  & -00d12m27.s1  & 34.1 & C  & 111210  & 5062309  & 3C286 & OQ208 & J1557-0001   \\
	&		 &		 &      & L  & 111230  & 4812476  & 3C286 & OQ208 & J1557-0001   \\
\enddata
%% Text for table notes should follow after the \enddata but before
%% the \end{deluxetable}. Make sure there is at least one \tablenotemark
%% in the table for each \tablenotetext.
\tablecomments{Observations of the galaxies, with the date, scheduling block identification number and primary, secondary and zero polarization leakage calibrators}
\tablenotetext{a}{Large galaxies denoted with an asterisk were observed in two pointings in C-band}
\tablenotetext{b}{Updated distances (see Sec.~\ref{sec:distances}) derived with TGRB are shown with an asterisk. Virgo cluster galaxies are indicated with a V}
\tablenotetext{c}{Alternate names for primary and leakage calibrators: 3C84 = J0319+4130, 3C286 = J1331+305, 3C48 = OQ208 = QSO B1404+2841 or J1407+2827}
\end{deluxetable}

%%%%%%%%%%%%% TABLE 2 - Two pointings coordinates %%%%%%%%%%%%%%%

\clearpage
\begin{deluxetable}{ccccc}
\tabletypesize{\scriptsize}
%\rotate
\tablecaption{Two pointings of large galaxies \label{tab:twopointings}}
\tablewidth{0pt}
\tablehead{
\colhead{Galaxy} & \colhead{RA 1} & \colhead{Dec 1} & \colhead{RA 2} & \colhead{Dec 2}\\
(1)&(2)&(3)&(4)& (5)\\
}
\startdata
N891  &02h 22m 37.21s & 42d 22m 41.2s    & 02h 22m 29.61s & 42d 19m 12.6s  \\
N3628 &11h 20m 24.50s & 13d 34m 55.7s    & 11h 20m 9.52s  & 13d 35m 50.1s  \\
N4244 &12h 17m 36.71s & 37d 49m 40.9s    & 12h 17m 22.61s & 37d 47m 10.3s  \\
N4565 &12h 36m 26.58s & 25d 57m 54.7s    & 12h 36m 14.98s & 26d 00m 36.6s  \\
N4631 &12h 42m 16.88s & 32d 32m 37.2s    & 12h 41m 59.14s & 32d 32m 21.6s  \\
N4594 &12h 40m 07.09s & - 11d 37m 23.0s  & 12h 39m 51.77s & - 11d 37m 23.0s\\
N5084 &13h 20m 24.88s & - 21d 49m 19.8s  & 13h 20m 8.96s  & - 21d 49m 58.8s\\
N5907 &15h 15m 59.49s & 56d 18m 1.6s     & 15h 15m 48.05  & 56d 21m 25.5s  \\
\enddata
\tablecomments{Column 1: Galaxy name; Column 2: Right Ascension of pointing 1; Column 3: Declination of pointing 1; Column 4: Right
Ascension of pointing 2; Column 5: Declination of pointing 2.}
\end{deluxetable}

%% If you use the table environment, please indicate horizontal rules using
%% Do not put multiple tabular environments within a single table.
%% The optional \label should appear inside the \caption command.

%% If the table is more than one page long, the width of the table can vary
%% from page to page when the default \tablewidth is used, as below.  The
%% individual table widths for each page will be written to the log file; a
%% maximum tablewidth for the table can be computed from these values.
%% The \tablewidth argument can then be reset and the file reprocessed, so
%% that the table is of uniform width throughout. Try getting the widths
%% from the log file and changing the \tablewidth parameter to see how
%% adjusting this value affects table formatting.

%% The \dataset{} macro has also been applied to a few of the objects to
%% show how many observations can be tagged in a table.

%%%%%%%%%%%%% TABLE 3 - Peeling coordinates for 2 galaxies %%%%%%%%%%%%%%%

%\clearpage
\begin{deluxetable}{cccccc}
\tabletypesize{\scriptsize}
%\rotate
\tablecaption{Peeled sources in the L-band fields of two galaxies  \label{tab:peeling}}
\tablewidth{0pt}
\tablehead{
\colhead{Galaxy} & \colhead{Source name} & \colhead{Angular distance}& \colhead{Intensity before/after} & \colhead{RA} & \colhead{Dec}\\
   &   & (\arcmin)  & (mJy/beam) & & \\
(1)&(2)&(3)&(4)&(5)&(6)\\
}
\startdata
N660  &  -  & 15  & 95/26 & 01h 42m 18.8s & +13d 27m 48.6s  \\
N4388 & M87 & 75  & 132/22 & 12h 30m 48s   & +12d 23m 30.2s  \\
\enddata
\tablecomments{Column 1: Galaxy name; Column 2: Name of interfering source 1; Column 3: distance in arcminutes between galaxy and interfering source; Column 4: Intensity of interfering source before/after peeling ; Column 5: Right
Ascension; Column 6: Declination.}
\end{deluxetable}

%%%%%%%%%%%%% TABLE 4 - Imaging parameters C-band %%%%%%%%%%%%%%%

\clearpage
\begin{deluxetable}{ccccccccccc}
\tabletypesize{\scriptsize}
%\rotate
\tablecaption{Imaging results C-band \label{tab:imagingparametersC}}
\tablewidth{0pt}
\tablehead{
\colhead{Galaxy} & \colhead{weighting} & \colhead{$B_{maj}$} & \colhead{$B_{min}$} & \colhead{$BPA$} & \colhead{$\nu_0$} & \colhead{rms I} &\colhead{Map peak} & \colhead{DR} & \colhead{rms QU} & \colhead{pol. peak} \\
&&["]&["]&[degree]& [GHz]& [$\mu$Jy/beam]& [mJy/beam]& & [$\mu$Jy/beam] & [$\mu$Jy/beam] \\
(1)&(2)&(3)&(4)&(5)&(6)&(7)&(8)&(9)&(10)&(11)\\
}
\startdata
N660  &  rob 0 & 9.55  & 9.14  & -18.72  &5.99904 & 8.6  & 603         & 70116    & 6.0	 &  369.1  \\
      &  uvtap & 16.13 & 16    & -58.32  &	  & 18   & 614         & 34111    & 7.0	 &  554.6 \\
N891  &  rob 0 & 9     & 8.81  & -79.46  &5.99932 & 6.5  & 7.21        & 1109	  & 6.6  &  115.0  \\
      &  uvtap & 15.31 & 14.63 & 88.05   &	  & 6.9  & 6.9         & 1000	  & 7.2  &  225.0 \\
N2613 &  rob 0 & 15.42 & 8.3   & -11.05  &5.99900 & 10.1 & 0.643       & 64	  & 10.4 &  57.1 \\
      &  uvtap & 19.94 & 15.82 & -11.44  &	  & 11.8 & 1.09        & 92	  & 11.5 &  63.6 \\
N2683 &  rob 0 & 9.36  & 8.76  & 11.71   &5.99900 & 9.1  & 3.27        & 359	  & 8.8  &  50.2 \\
      &  uvtap & 15.88 & 14.81 & 46.67   &	  & 9.4  & 3.31        & 352	  & 8.9  &  64.6 \\
N2820 &  rob 0 & 9.84  & 8.91  & -3.81   &5.99900 & 5.3  & 1.699       & 321	  & 5.3  &  48.7 \\
      &  uvtap & 16.18 & 15.36 & -3.884  &	  & 5.2  & 3.173       & 610	  & 5.3  &  97.3 \\
N2992 &  rob 0 & 14.33 & 8.84  & -13.3   &5.99900 & 8.4  & 65.6        & 7810	  & 8.0  &  151.7 \\
      &  uvtap & 18.51 & 16.76 & -14.26  &	  & 9.5  & 71.7        & 7547	  & 8.2  &  202.8 \\
N3003 &  rob 0 & 9.37  & 8.93  & -67.02  &5.99900 & 9.4  & 1.35        & 144	  & 8.6  &  41.1 \\
      &  uvtap & 13.78 & 13.61 & 16.65   &	  & 6.0	 & 8.11        & 1352	  & 9.2	 &  40.0 \\
N3044 &  rob 0 & 10.68 & 9.59  & -5.08   &5.99900 & 5.1  & 5.92        & 1161	  & 7.0	 &  102.9  \\
      &  uvtap & 13.78 & 13.61 & 16.65   &	  & 6.0	 & 8.11        & 1352	  & 7.0	 &  138.4  \\
N3079 &  rob 0 & 9.32  & 8.72  & -13.89  &5.99900 & 6.5  & 202.07      & 31088    & 5.9  &  1987.1 \\
      &  uvtap & 16.16 & 15.44 & -9.5    &	  & 8.0	 & 221.93      & 27741    & 5.8  &  2759.5 \\
N3432 &  rob 0 & 11.94 & 9.97  & 3.43    &5.99900 & 7.2  & 2.33        & 324	  & 7.1  &  39.2 \\
      &  uvtap & 15    & 13.14 & -179.3  &	  & 6.8  & 2.62        & 385	  & 6.5  &  49.3 \\
N3448 &  rob 0 & 9.1   & 8.6   & 3.4     &5.99900 & 5.6  & 5.36        & 957	  & 5.4  &  66.0 \\
      &  uvtap & 16    & 15.7  & -167.6  &	  & 6.6  & 8.46        & 1282	  & 5.9  &  105.5 \\
N3556 &  rob 0 & 9.2   & 8.7   & 7.6     &5.99900 & 5.3  & 3.45        & 651	  & 5.3  &  66.8 \\
      &  uvtap & 16    & 15.   & 14.1    &	  & 6.0	 & 6.9         & 1150	  & 5.7  &  91.9 \\
N3628 &  rob 0 & 9.62  & 9.47  & -31.09  &5.99842 & 9.5  & 66.96       & 7048	  & 10.2 &  174.2  \\
      &  uvtap & 16.83 & 15.15 & 75.57   &	  & 9.5  & 79.09       & 8325	  & 9.6  &  128.4 \\
N3735 &  rob 0 & 10.1  & 8.8   & -2.3    &5.99900 & 5.8  & 2.9         & 500	  & 5.6  &  109.5  \\
      &  uvtap & 16.4  & 15.6  & -5	 &	  & 6.2  & 5.3         & 855	  & 5.6  &  220.4 \\
N3877 &  rob 0 & 8.85  & 8.66  & 39.45   &5.99900 & 6.5  & 1.43        & 220	  & 6.4  &  25.6 \\
      &  uvtap & 16.07 & 15.27 & 22.61   &	  & 6.7  & 1.94        & 290	  & 6.6  &  33.7 \\
N4013 &  rob 0 & 9.07  & 8.88  & -31.87  &5.99900 & 6.4  & 3.88        & 606	  & 6.4  &  41.5 \\
      &  uvtap & 16.06 & 15.25 & -7.25   &	  & 6.6  & 4.58        & 694	  & 6.5  &  62.5 \\
N4096 &  rob 0 & 9     & 8.85  & -59.54  &5.99900 & 6.3  & 0.529       & 84	  & 6.3  &  26.9 \\
      &  uvtap & 15.85 & 15.32 & -16.41  &	  & 6.8  & 1.1         & 162	  & 6.6  &  28.7 \\
N4157 &  rob 0 & 9.1   & 8.74  & 44.64   &5.99900 & 6.3  & 1.47        & 3411     & 6.0  &  58.3 \\
      &  uvtap & 15.9  & 14.99 & 21.61   &	  & 6.2  & 3.6         & 581	  & 6.1  &  123.3 \\
N4192 &  rob 0 & 9.17  & 8.99  & -11.55  &5.99900 & 6.0	 & 3.807       & 635	  & 6.0  &  150.2  \\
      &  uvtap & 15.66 & 15    & 83.71   &	  & 6.0	 & 5.128       & 855	  & 5.9  &  222.2 \\
N4217 &  rob 0 & 8.91  & 8.73  & 29.41   &5.99900 & 6.3  & 2.63        & 417	  & 6.3  &  69.4 \\
      &  uvtap & 16.06 & 15.23 & 17.5    &	  & 6.3  & 4.52        & 717	  & 6.2  &  137.5 \\
N4244 &  rob 0 & 9.25  & 8.91  & -4.37   &5.99838 & 5.6  & 0.693       & 235      & 5.9  &  14.3 \\
      &  uvtap & 15.83 & 15.07 & -1.57   &	  & 5.8  & 0.84/1.39   & 240	  & 5.7  &  18.6 \\
N4302 &  rob 0 & 9.96  & 8.96  & -21.41  &5.99900 & 15.5 & 1.29        & 146	  & 15.2 &  73.1 \\
      &  uvtap & 18.7  & 18.4  & 32.98   &	  & 19.5 & 1.93        & 99	  & 17.0 &  113.3 \\
N4388 &  rob 0 & 9.9   & 8.99  & -24.75  &5.99900 & 13.9 & 18.1        & 1302	  & 13.5 &  436.1  \\
      &  uvtap & 16.2  & 15.97 & -23.7   &	  & 18.0 & 25.17       & 1398	  & 16.4 &  462.7 \\
N4438 &  rob 0 & 9.83  & 9.17  & -19.86  &5.99900 & 16.0 & 22.91       & 1432	  & 15.0 &  221.8  \\
      &  uvtap n/a&       &       & 	 &	  &	 &	       &	  &	 &   \\
N4565 &  rob 0 & 9.02  & 8.82  & 85.48   &5.99834 & 7.4  & 2.64        & 357	  & 7.4  &  49.7 \\
      &  uvtap & 15.96 & 14.83 & 42.18   &	  & 8.0	 & 3.11        & 389	  & 7.8  &  105.0 \\
N4594 &  rob 0 & 13.32 & 8.91  & 1.1     &5.99841 & 12.9 & 102         & 7907	  & 11.4 &  228.2 \\
      &  uvtap & 18.09 & 16.07 & 0.8	 &	  & 14.4 & 103.3       & 7159	  & 13.7 &  245.0 \\
N4631 &  rob 0 & 8.88  & 8.57  & -6.88   &5.99835 & 7.7  & 6.8         & 883	  & 6.9  &  122.4 \\
      &  uvtap & 15.71 & 14.73 & 86.79   &	  & 9.2  & 13.9        & 1511	  & 9.6  &  368.6 \\
N4666 &  rob 0 & 11.06 & 9.5   & -179.82 &5.99900 & 12.5 & 7	       & 560	  & 12.0 &  210.8  \\
      &  uvtap & 14.1  & 13.54 & -176.99 &	  & 12.5 & 10.61       & 849	  & 11.5 &  332.0 \\
N4845 &  rob 0 & 10.98 & 9.06  & -1.4    &5.99900 & 15.0 & 424.7       & 28313    & 14.5 &  337.3 \\
      &  uvtap n/a &       &       & 	 &	  &	 &	       &	  &	 &   \\
N5084 &  rob 0 & 15.64 & 8.35  & -5.76   &5.99841 & 7.9  & 29.1        & 3684	  & 8.0  &  76.4 \\
      &  uvtap & 20.33 & 16.23 & -11.18  &	  & 8.6  & 29.4        & 3419	  & 9.4  &  83.3 \\
N5297 &  rob 0 & 9.03  & 8.78  & 35.28   &5.99900 & 6.3  & 0.267       & 42	  & 5.6  &  24.9 \\
      &  uvtap & 15.66 & 14.87 & 21.77   &	  & 6.1  & 0.485       & 80	  & 6.1  &  29.8 \\
N5775 &  rob 0 & 10.08 & 9.34  & -22.79  &5.99900 & 5.0	 & 4.45        & 890	  & 5.0  &  104.0  \\
      &  uvtap & 16.1  & 14.96 & 75.53   &	  & 5.8  & 7.77        & 1340	  & 5.3  &  214.2 \\
N5792 &  rob 0 & 10.34 & 9.18  & -12.72  &5.99900 & 5.3  & 6.183       & 1167	  & 5.3  &  73.9 \\
      &  uvtap & 16.09 & 15.54 & -87.29  &	  & 5.4  & 9.29        & 1720	  & 5.4  &  76.9 \\
N5907 &  rob 0 & 9.78  & 8.08  & 16.32   &5.99847 & 9.3  & 0.875/5.76 & 623      & 9.1  &  140.7 \\
      &  uvtap & 16.65 & 14.55 & 65.68   &	  & 9.8  & 1.17/5.22   & 533	  & 9.6  &  115.5 \\
U10288\tablenotemark{a}&  rob 0 & 10.96 & 9.46  & -28.47  &5.99900 & 7.0	 & 8.94        & 1277	  & 5.4  &  146.6 \\
      &  uvtap & 13.99 & 13.42 & -59.75  &	  & 8.5  & 9.39        & 1105	  & 5.4  &  169.8 \\
\enddata
%% Text for table notes should follow after the \enddata but before
%% the \end{deluxetable}. Make sure there is at least one \tablenotemark
%% in the table for each \tablenotetext.
\tablecomments{Col. 1: Galaxy name; Col. 2: Weighting, where "rob 0" refers to Briggs weighting with robust value set to 0, "uvtap" refers to a uv-tapered version of the former; Columns 3-5: synthesized beam parameters; Col. 6: C-band central frequency in GHz (varying due to differences in flagging); Col. 7: Stokes I rms noise; Col. 8: Peak intensity of the galaxy. In the case of two values given, the first value is the peak intensity of the galaxy and the second higher value is the peak of the map when this occurs outside of the galaxy; Col. 9: Dynamic range in image (map peak intensity over noise); Col. 10: Stokes Q and U average rms noise; Col. 11: Peak intensity of the polarization map (measured from non-PB-corrected maps). }
\tablenotetext{a}{Map peak and polarization peak values given for UGC~10288 are of the background source.}
\end{deluxetable}

%%%%%%%%%%%%% TABLE 5 - Imaging parameters L-band %%%%%%%%%%%%%%%

\clearpage
\begin{deluxetable}{ccccccccccc}
\tabletypesize{\scriptsize}
%\rotate
\tablecaption{Imaging results L-band \label{tab:imagingparametersL}}
\tablewidth{0pt}
\tablehead{
\colhead{Galaxy} & \colhead{weighting} & \colhead{$B_{maj}$} & \colhead{$B_{min}$} & \colhead{$BPA$} & \colhead{$\nu_0$} &\colhead{rms I} &\colhead{Map peak} & \colhead{DR} & \colhead{rms QU} & \colhead{pol. peak} \\
&&["]&["]&[degree]& [GHz]&[$\mu$Jy/beam]& [mJy/beam]& & [$\mu$Jy/beam] & [$\mu$Jy/beam] \\
(1)&(2)&(3)&(4)&(5)&(6)&(7)&(8)&(9)&(10)&(11)\\
}
\startdata
N660  &  rob 0    & 39.78 & 36.27 & 80.22  &  1.57502 &  60  & 422.5	  & 7042       & 28	 & 191 \\ 
      & uvtap     & 49.85 & 47.29 & -86.89 & 	      &  75  & 441.8	  & 5891       & 28	 & 177 \\ 
N891  &  rob 0    & 36.42 & 32.48 & -74.29 &  1.57498 &  60  & 73.4/78.4  & 1307       & 40\tablenotemark{a}	 & 455 \\
      & uvtap     & 44.94 & 43.11 & 67.61  & 	      &  95  & 96.7	  & 1018       & 44	 & 647 \\
N2613 &  rob 0    & 73.43 & 41.97 & -29.89 &  1.57497 &  42  & 11.96/17.2 & 410        & 35	 & 176 \\
      & uvtap     & 76.55 & 51.18 & -32.11 & 	      &  45  & 13.4/16.9  & 376        & 40	 & 191 \\
N2683 &  rob 0    & 33.9  & 31.31 & -43.8  &  1.57481 &  30  & 8.6	  & 287        & 27	 & 223 \\
      & uvtap     & 44.45 & 41.69 & -44.23 & 	      &  35  & 11.7	  & 334        & 27	 & 218 \\
N2820 &  rob 0    & 34.8  & 30.56 & -32.93 &  1.57479 &  40  & 18.54/39.9 & 998        & 32	 & 132 \\
      & uvtap     & 43.97 & 39.88 & -36.87 & 	      &  40  & 23.6	  & 590        & 32	 & 152 \\
N2992 &  rob 0    & 52.08 & 33.1  & -13.55 &  1.57481 &  30  & 193.2	  & 6440       & 26	 & 139 \\
      & uvtap     & 59.34 & 45.81 & -20.87 & 	      &  32  & 195.4	  & 6106       & 27	 & 167 \\
N3003 &  rob 0    & 33.86 & 30.7  & -46.95 &  1.57478 &  30  & 8.93	  & 298        & 27	 & 102 \\
      & uvtap     & 44.64 & 40.58 & -47.84 & 	      &  29  & 11.5	  & 397        & 27	 & 92 \\
N3044 &  rob 0    & 41.79 & 32.25 & -48.11 &  1.57479 &  28  & 43.96	  & 1570       & 25	 & 246 \\
      & uvtap     & 50.92 & 42.39 & -61.13 & 	      &  27  & 20.76	  & 769        & 27	 & 270 \\
N3079 &  rob 0    & 34.88 & 31.49 & -48.61 &  1.57477 &  45  & 427.3	  & 9496       & 30	 & 488 \\
      & uvtap     & 45.64 & 41.61 & -48.18 & 	      &  45  & 481.2	  & 10693      & 30	 & 646 \\
N3432 &  rob 0    & 32.62 & 31.82 & -64.03 &  1.57474 &  36  & 12.16/27.83& 773        & 32	 & 181 \\
      & uvtap     & 42.34 & 41.43 & -51.85 & 	      &  40  & 16.8/27    & 675        & 32	 & 195 \\
N3448 &  rob 0    & 34.55 & 31.4  & -23.24 &  1.57475 &  35  & 25.46	  & 727        & 26	 & 114 \\
      & uvtap n/a &	  &	  &        & 	      &      &  	  &	       &	 &  \\
N3556 &  rob 0    & 34.45 & 31.31 & -22.75 &  1.57475 &  38  & 33.87	  & 891        & 25	 & 372 \\
      & uvtap     & 45.91 & 41.74 & -21.92 & 	      &  42  & 48.2	  & 1148       & 26	 & 503 \\
N3628 &  rob 0    & 32.27 & 32.65 & -37.4  &  1.57473 &  49  & 249	  & 5082       & 30	 & 30 \\ 
      & uvtap     & 47.19 & 42.47 & -41.85 & 	      &  55  & 267	  & 4855       & 28	 & 40 \\ 
N3735 &  rob 0    & 36.8  & 32.1  & -15.7  &  1.57477 &  32  & 33.5	  & 1047       & 25	 & 213 \\ 
      & uvtap     & 46.6  & 41.1  & -21.7  & 	      &  31  & 42	  & 1355       & 25	 & 250 \\
N3877 &  rob 0    & 34.12 & 32.61 & -31.6  &  1.57473 &  30  & 7.93	  & 264        & 27	 & 180 \\ 
      & uvtap     & 45.41 & 41.36 & -27.75 & 	      &  37  & 10.2	  & 276        & 25	 & 188 \\
N4013 &  rob 0    & 34.15 & 31.54 & -65.96 &  1.57472 &  32  & 13.8	  & 431        & 27	 & 103 \\
      & uvtap     & 44.98 & 41.45 & -64.79 & 	      &  38  & 15.5/24.6  & 647        & 26	 & 116 \\
N4096 &  rob 0    & 33.85 & 31.73 & -35.65 &  1.57473 &  30  & 7.099	  & 237        & 27	 & 133 \\
      & uvtap     & 44.75 & 41.51 & -34.28 & 	      &  32  & 10.33	  & 323        & 26	 & 180 \\
N4157 &  rob 0    & 34.12 & 31.65 & -23.32 &  1.57488 &  29  & 31.42/53.36& 1840       & 28	 & 254 \\
      & uvtap     & 45.19 & 41.48 & -21.07 & 	      &  35  & 43.8/54.5  & 1557       & 27	 & 291 \\
N4192 &  rob 0    & 35.52 & 32.66 & -52.17 &  1.57471 &  40  & 17.5	  & 438        & 28	 & 205 \\
      & uvtap n/a &	  &	  &    	   & 	      &      &  	  &	       &	 &  \\
N4217 &  rob 0    & 34.01 & 31.46 & -29.04 &  1.57472 &  28  & 24	  & 857	       & 27	 & 264\\
      & uvtap     & 45.31 & 41.07 & -26.28 & 	      &  30.4& 33.1	  & 1089       & 28	 & 368\\
N4244 &  rob 0    & 33.98 & 32.46 & -47.1  &  1.57471 &  30  & 2.18/9.87  & 329        & 27	 & 114\\
      & uvtap     & 45.43 & 41.84 & -43.9  & 	      &  30  & 10.1	  & 337	       & 27	 & 93\\
N4302 &  rob 0    & 35.82 & 33.55 & -70.51 &  1.57470 &  46  & 9.73/15.32 & 333	       & 35	 & 319\\
      & uvtap     & 47    & 42.8  & -81.78 & 	      &  60  & 15.11	  & 252	       & 35	 & 409\\
N4388 &  rob 0    & 36.32 & 33.24 & -48.92 &  1.59888 &  150 & 79.92/296.2& 1975       & 65	 & 436\\
      & uvtap n/a &	  &	  &        & 	      &      &            &	       &	 &  \\
N4438 &  rob 0    & 34.7  & 30.21 & -53.59 &  1.77468 &  130 & 79.2	  & 609	       & 50      & 237\\
      & uvtap n/a &	  &	  &        & 	      &      &            &	       &	     &  \\
N4565 &  rob 0    & 34.5  & 32.32 & -89.06 &  1.57470 &  30  & 9.36/48    & 1600       & 27	 & 270\\
      & uvtap     & 43.27 & 41.92 & -83.71 & 	      &  32  &12.58/48.9  & 1528       & 24	 & 451\\
N4594 &  rob 0    & 47.89 & 32.62 & -4.6   &  1.57471 &  31  & 81.4	  & 2626       & 25	 & 209\\
      & uvtap     & 54.41 & 44.46 & -8.23  & 	      &  32  & 81.9	  & 2559       & 25	 & 254\\
N4631 &  rob 0    & 35    & 32.4  & -89.34 &  1.57488 &  31  & 90.1	  & 2906       & 28	 & 353 \\
      & uvtap     & 32.81 & 41.72 & -82.91 & 	      &  31  & 123	  & 3955       & 33	 & 449 \\
N4666 &  rob 0    & 37.4  & 36.08 & -30.42 &  1.57470 &  23  & 102.4	  & 4452       & 25	 & 700\\
      & uvtap     & 47.95 & 44.7  & -72.66 & 	      &  27  & 132	  & 4981       & 25	 & 865\\
N4845 &  rob 0    & 38.58 & 34.27 & -5.22  &  1.57470 &  40  & 224.8	  & 5620       & 27	 & 556\\
      & uvtap     &	  &	  &	   & 	      &      &  	  &	       &	 & \\
N5084 &  rob 0    & 57.12 & 31.48 & -5.85  &  1.57472 &  33  & 35.6	  & 1079       & 29	 & 395\\
      & uvtap     & 65.86 & 43.94 & -8.33  & 	      &  35  & 36.5	  & 1043       & 30	 & 426\\
N5297 &  rob 0    & 34.33 & 32.17 & 5.23   &  1.57471 &  34  & 4.25	  & 125	       & 27	 & 138\\
      & uvtap     & 45.16 & 41.38 & -5.49  & 	      &  32  & 6.11	  & 191	       & 26	 & 143\\
N5775 &  rob 0    & 40.35 & 35.4  & -42.84 &  1.57468 &  35  & 57.5	  & 1643       & 26	 & 542\\
      & uvtap     & 51.16 & 45.04 & -63.19 & 	      &  33  & 75.5	  & 2323       & 27	 & 645\\
N5792 &  rob 0    & 40.13 & 35.23 & -29.48 &  1.57469 &  32  & 31.92	  & 998	       & 25	 & 1366\\
      & uvtap     & 49.54 & 47.71 & -39.27 & 	      &  35  & 34.4	  & 983	       & 25	 & 1461\\
N5907 &  rob 0    & 32.99 & 30.67 & 44.87  &  1.57474 &  30  & 9.45/52.79 & 1759       & 28	 & 556\\
      & uvtap     & 42.49 & 39.93 & 42.4   & 	      &  30  & 13.4/31.8  & 1060       & 28	 & 603\\
U10288\tablenotemark{b}&  rob 0    & 40.23 & 34.34 & -31.32 &  1.57488 &  39  & 98.7	  & 2531       & 35      & 3027\\
      & uvtap n/a &	  &	  &	   & 	      &      &  	  &	       &	 &  \\
\enddata
\tablecomments{Col. 1: Galaxy name; Col. 2: Weighting, where "rob 0" denotes Briggs weighting with robust value set to 0, "uvtap" refers to a uv-tapered version of the former; Columns 3-5: synthesized beam parameters; Col. 6: L-band central frequency in GHz (varying due to differences in flagging); Col. 7: Stokes I rms noise; Col. 8: Peak intensity of the galaxy. In the case of two values given, the first value is the peak intensity of the galaxy and the second higher value is the peak of the map; Col. 9: Dynamic range in image; Col. 10: Stokes Q and U average rms noise; Col. 11: Peak intensity of the polarization map (measured from non-PB-corrected maps). }
\tablenotetext{a}{A previous observation (an extra 10 minutes on source) was included in the L-band Stokes Q and U imaging to increase sensitivity. It was however excluded from Stokes I imaging due to artifact contamination.}
\tablenotetext{b}{Map peak and polarization peak values given for UGC~10288 are of the background source.}
\end{deluxetable}

\clearpage 
\begin{deluxetable}{cccccccc}
\tabletypesize{\scriptsize}
%\rotate
\tablecaption{Star formation rates \label{tab:sfr_etc}}
\tablewidth{0pt}
\tablehead{
\colhead{Galaxy} & \colhead{Flux density} & \colhead{Flux} & \colhead{Diameter} &\colhead{Diameter} &\colhead{SFR} & \colhead{SFR surface density} & \colhead{Uncertainty} \\
& [Jy] &[$\times 10^{-13} W/m^2$] & [\arcmin]&[kpc] &[{$M_\sun$}/yr]& [$\times 10^{-3} M_{\sun}/yr/kpc^2$] & \\
(1)&(2)&(3)&(4)&(5)&(6)&(7)&(8)\\
}
\startdata
N660   &  5.70 & 7.77 & 3.02 & 10.8 &  2.74    & 29.96   & 0.02    \\
N891   &  5.90 & 8.04 & 9.50 & 25.1 &  1.55    & 3.13    & 0.02    \\
N2613  &  0.99 & 1.35 & 4.95 & 33.7 &  1.73    & 1.94    & 0.02    \\
N2683  &  0.74 & 1.01 & 5.15 & 9.39 &  0.09    & 1.34    & 0.02    \\
N2820  &  0.28 & 3.80 & 1.78 & 13.8 &  0.62    & 4.20    & 0.02    \\
N2992  &  0.88 & 1.19 & 1.27 & 12.5 &  3.22    & 26.13   & 0.03    \\
N3003  &  0.33 & 4.46 & 3.75 & 27.7 &  0.67    & 1.11    & 0.02    \\
N3044  &  0.73 & 9.91 & 3.07 & 18.1 &  0.95    & 3.70    & 0.02    \\
N3079  &  2.56 & 3.50 & 4.33 & 26.0 &  3.46    & 6.54    & 0.02    \\
N3432  &  0.54 & 7.42 & 3.63 & 9.96 &  0.15    & 1.97    & 0.02    \\
N3448  &  0.48 & 6.59 & 1.75 & 12.5 &  0.92    & 7.55    & 0.02    \\
N3556  &  3.43 & 4.68 & 6.08 & 24.9 &  2.17    & 4.44    & 0.02    \\
N3628  &  4.41 & 6.02 & 9.90 & 24.5 &  1.01    & 2.15    & 0.02    \\
N3735  &  0.84 & 1.14 & 2.82 & 34.4 &  1.10\tablenotemark{a}    & 1.2\tablenotemark{a}    & 0.02    \\
N3877  &  0.92 & 1.25 & 3.58 & 18.5 &  0.92    & 3.43    & 0.02    \\
N4013  &  0.59 & 7.98 & 3.45 & 16.1 &  0.48    & 2.35    & 0.02    \\
N4096  &  0.79 & 1.08 & 3.92 & 11.8 &  0.27    & 2.46    & 0.02    \\
N4157  &  1.62 & 2.21 & 3.67 & 16.6 &  1.25    & 5.77    & 0.02    \\
N4192  &  0.96 & 1.31 & 6.20 & 24.4 &  0.56    & 1.19    & 0.02    \\
N4217  &  1.14 & 1.55 & 3.90 & 23.4 &  1.53    & 3.57    & 0.02    \\
N4244  &  0.38 & 5.22 & 11.53 & 14.8 &  0.02    & 0.14    & 0.03    \\
N4302  &  0.44 & 6.00 & 3.82 & 21.6 &  0.53    & 1.45    & 0.02    \\
N4388  &  2.18 & 2.97 & 2.30 & 11.1 &  0.07\tablenotemark{b}    & 0.72\tablenotemark{b}   & 0.02    \\
N4438  &  0.19 & 2.64 & 1.32 & 3.98 &  0.07    & 5.35    & 0.03    \\
N4565  &  1.65 & 2.25 & 10.43 & 36.1 &  0.74    & 0.73    & 0.02    \\
N4594  &  0.62 & 8.44 & 5.97 & 22.0 &  0.32    & 0.83    & 0.02    \\
N4631  &  7.62 & 1.04 & 10.85 & 23.4 &  1.33    & 3.10    & 0.02    \\
N4666  &  3.03 & 4.13 & 4.03 & 32.3 &  7.29    & 8.92    & 0.02    \\
N4845  &  0.52 & 7.14 & 2.10 & 10.4 &  0.48    & 5.68    & 0.02    \\
N5084  &  0.06 & 7.73 & 0.73 & 4.99 &  0.10    & 5.04    & 0.05    \\
N5297  &  0.24 & 3.33 & 2.20 & 25.9 &  1.27    & 2.41    & 0.02    \\
N5775  &  1.99 & 2.71 & 3.80 & 32.0 &  5.28    & 6.58    & 0.02    \\
N5792  &  0.82 & 1.12 & 2.57 & 23.7 &  2.63    & 5.98    & 0.02    \\
N5907  &  1.73 & 2.36 & 7.35 & 35.9 &  1.56    & 1.54    & 0.02    \\
U10288 &  0.11 & 1.51 & 2.15 & 21.3 &  0.41    & 1.15    & 0.03    \\
\enddata
%% Text for table notes should follow after the \enddata but before
%% the \end{deluxetable}. Make sure there is at least one \tablenotemark
%% in the table for each \tablenotetext.
\tablecomments{Col. 1: Name; Col. 2 and 3: 22 \micron\ flux, given in two different units (The values in Col. 3 were used for Fig.~\ref{fig:spitzerwise}); Col. 4 and 5: Angular diameters; Col.6: Star formation rates in solar masses/year; Col 7: Star formation rate density $\times10^{-3}$ in solar masses/year and $kpc^2$; Col. 8: Fractional error on surface density.}
\tablenotetext{a}{The SFR and SFR surface density values of NGC~3735 were adjusted from 4.71 and 0.005 respectively, to the values given here (lower limits), in order to account for the central AGN (see Sec.~\ref{sec:AGNcontamination})}
\tablenotetext{b}{The SFR and SFR surface density values of NGC~4388 were adjusted from 1.91 and 0.020 respectively, to the values given here (lower limits), in order to account for the central AGN (see Sec.~\ref{sec:AGNcontamination})}
\end{deluxetable}

%%%%%%%%%%%%% TABLE 7 - flux densities,  %%%%%%%%%%%%%%%
\clearpage 
\begin{deluxetable}{ccccc}
\tabletypesize{\scriptsize}
%\rotate
\tablecaption{Flux densities \label{tab:fluxdensities}}
\tablewidth{0pt}
\tablehead{
\colhead{Galaxy} & \colhead{Flux density C} & \colhead{Flux density C uncertainty} & \colhead{Flux density L} &\colhead{Flux density L uncertainty}\\
& & & &\\
(1)&(2)&(3)&(4)&(5)\\
}
\startdata
N660   & 657.9   & 13.2    & 525.4   &  10.5 \\
N891   & 208.7   & 6.9\tablenotemark{a}    & 743.9   &  14.9 \\
N2613  & 15.3	 & 0.3     & 59.6    &  2.4\tablenotemark{a} \\
N2683  & 20.3	 & 0.8\tablenotemark{a}    & 66.6    &  6.5\tablenotemark{a} \\
N2820  & 19.1	 & 0.4     & 61.8    &  1.2  \\
N2992  & 80.4	 & 1.6     & 204.9   &  4.1  \\
N3003  & 10.8	 & 0.5\tablenotemark{a}    & 34.9    &  0.7  \\
N3044  & 37.5	 & 0.9\tablenotemark{a}    & 104.2   &  2.1  \\
N3079  & 365.4   & 7.3     & 811.0   &  16.2 \\
N3432  & 26.3	 & 0.5     & 83.3    &  1.9\tablenotemark{a} \\
N3448  & 20.5	 & 0.4     & 46.0    &  0.9  \\
N3556  & 79.2	 & 4.7\tablenotemark{a}    & 291.5   &  5.8  \\
N3628  & 184.6   & 3.7     & 527.5   &  10.5 \\
N3735  & 24.9	 & 0.5     & 81.3    &  1.6  \\
N3877  & 12.9	 & 0.3\tablenotemark{a}    & 42.7    &  0.9  \\
N4013  & 12.6	 & 0.3     & 37.9    &  0.8\tablenotemark{a} \\
N4096  & 16.3	 & 0.3     & 57.1    &  1.1  \\
N4157  & 55.1	 & 1.1     & 184.5   &  3.7  \\
N4192  & 24.4	 & 0.5     & 80.6    &  1.6  \\
N4217  & 35.4	 & 0.7     & 111.5   &  2.2  \\
N4244  & 9.0	 & 0.6\tablenotemark{a}    & 18.1    &  0.6\tablenotemark{a} \\
N4302  & 12.0	 & 0.3\tablenotemark{a}    & 45.1    &  0.9  \\
N4388  & 62.2	 & 1.2     & 130.9   &  2.6  \\
N4438  & 54.5	 & 1.1	   & 132.2   &  2.6  \\
N4565  & 42.3	 & 1.1\tablenotemark{a}    & 152.2   &  3.0  \\
N4594  & 128.3   & 2.6     & 93.7    &  1.9  \\
N4631  & 284.4   & 7.4\tablenotemark{a}    & 1083.0  &  37.0\tablenotemark{a}\\
N4666  & 125.3   & 2.5     & 404.5   &  8.1  \\
N4845  & 432	 & 8.6	   & 230.0   &  4.6  \\
N5084  & 36.1	 & 1.0\tablenotemark{a}    & 40.7    &  0.8  \\
N5297  & 6.7	 & 0.4\tablenotemark{a}    & 24.4    &  0.5  \\
N5775  & 74.4	 & 1.5     & 255.0   &  5.1  \\
N5792  & 20.2	 & 0.4     & 57.7    &  1.2  \\
N5907  & 51.5	 & 1.0     & 181.5   &  3.6  \\
U10288 & 1.53	 & 0.5\tablenotemark{b}    & 4.4     &  0.5\tablenotemark{b} \\
\enddata
%% Text for table notes should follow after the \enddata but before
%% the \end{deluxetable}. Make sure there is at least one \tablenotemark
%% in the table for each \tablenotetext.
\tablecomments{Col. 1: Name; Col. 2: Flux density C-band (mJy); Col. 3: Uncertainty of flux density C-band, in most cases 2\% calibration errors; Col. 4: Flux density L-band (mJy); Col. 5: Uncertainty of flux density L-band, in most cases 2\% calibration errors.}
\tablenotetext{a}{The error between the two weightings is larger than the 2\% calibration error, and in these cases, we use the former.} 
\tablenotetext{b}{For UGC~10288, we refer to Paper III.}
\end{deluxetable}

\end{document}